\documentclass[aip,reprint]{revtex4-1}
\usepackage{amsmath,graphicx,dcolumn,hyperref,xcolor}

\begin{document}

\title{Ion Channel Dynamics in Temperature-Dependent Weibel Instability Saturation}

\author{Vivek Shrivastav}
\thanks{Email: vivekshrivastav1998@gmail.com}
\affiliation{Department of Physics, Sikkim University, Gangtok, India, 737102}

\author{Mani K Chettri}
\affiliation{Department of Physics, Sikkim University, Gangtok, India, 737102}

\author{Hemam D Singh}
\affiliation{Department of Physics, Netaji Subhas University of Technology, New Delhi, India, 110078}

\author{Britan Singh}
\affiliation{Department of Physics, Sikkim University, Gangtok, India, 737102}

\author{Rupak Mukherjee}
\thanks{Corresponding author: rmukherjee@cus.ac.in}
\affiliation{Department of Physics, Sikkim University, Gangtok, India, 737102}

\begin{abstract}
We present 1X2V continuum Vlasov-Maxwell simulations of interpenetrating plasma beams with mobile ions. While the early-time evolution is similar to the stationary-ion case, the late-time dynamics are dominated by the ion-Weibel instability. As ion channels merge, the magnetic energy increases and the magnetic structures extend further along the beam direction. Electrons rapidly reach thermal equilibrium, whereas ions retain distinct bulk velocities for much longer and thermalize more slowly. These results are relevant to collisionless shock formation in astrophysical compact objects and laser-plasma experiments. Wind/SWE observations place all four simulated cases in the firehose/Weibel-unstable region of the proton temperature anisotropy diagram, and MMS1 observations of a quasi-perpendicular bow shock ($\theta_{Bn}\approx83^\circ$, $M_A\approx27$) show a qualitatively similar electron-ion thermalization disparity.
\end{abstract}

\maketitle

\section{Introduction}
\label{introduction}

\begin{figure}[h!]
\centering
\includegraphics[width=0.9\linewidth]{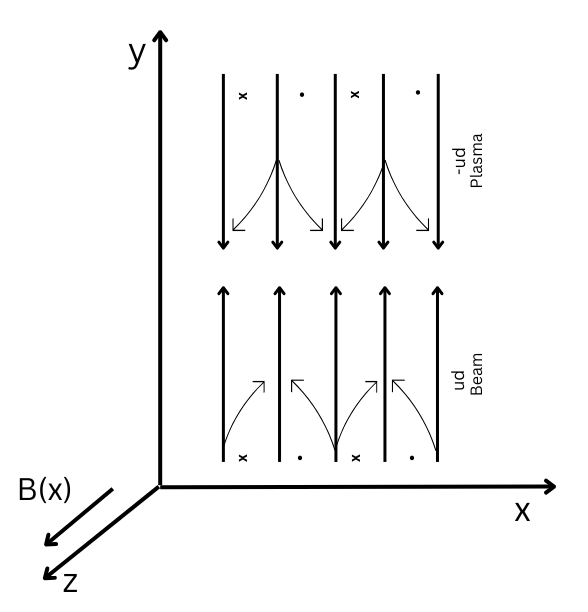}
\caption{Schematic diagram showing the mechanism of Weibel instability in a counter-streaming beam of plasma. The beam electrons (drifting upward with velocity $+u_d$) and the plasma electrons (drifting downward with velocity $-u_d$) experience opposing Lorentz forces in the presence of a transverse magnetic perturbation $B_z$. This causes them to concentrate into current filaments, which amplify the magnetic field, driving the instability.}
\label{paper WI}
\end{figure}

Weibel~\cite{weibel1959spontaneously} first observed that plasma having sufficiently anisotropic velocity distributions can sustain self-excited transverse electromagnetic waves. The Weibel instability has received significant attention in the laboratory and astrophysical plasma communities in recent years. Intergalactic space is filled with plasma and its presence has important consequences for the generation of magnetic fields at the cosmological scale. Medvedev \& Loeb~\cite{medvedev1999generation} showed through a linear kinetic treatment how Weibel instability can generate a quasi-static and strong magnetic field in colliding plasmas. A significant amount of energy from moving plasma particles is converted into magnetic energy due to electromagnetic instabilities. These instabilities play an important role in generating and enhancing magnetic fields. When plasma has temperature anisotropy, magnetization arises from the Weibel instability~\cite{skoutnev2019temperature,shrivastav2026weibel,hanebring2026weibel}. A thorough investigation into the growth and nonlinear saturation of the Weibel instability, which leads to magnetic trapping~\cite{davidson1972nonlinear} of charged particles, is required to understand many laboratory and astrophysical processes, such as collisionless shock formation in astrophysical compact objects and laser-plasma experiments. Below we briefly describe the basic principles of the Weibel instability and the growth and saturation of field energies. Let us consider a beam of electrons streaming through the plasma with drift velocity $\Vec{u}_d$. These beam electrons repel the background plasma electrons. The beam electrons are not deflected back because their directed momentum allows them to propagate through the background plasma, maintaining approximate current neutrality in the unperturbed state. The beam electrons will create a magnetic field. Consequently, the background plasma electrons will create a back electromotive force, generating a back current in accordance with Lenz's law. Hence, the plasma electron drift velocity can be considered $-\Vec{u}_d$. Initially unmagnetized and unperturbed, these beam and plasma currents cancel each other completely. However, even a small perturbation in the magnetic field will make the current finite. Now consider a situation where a magnetic perturbation $B_z\cos x\,\Hat{z}$ is applied. The schematic diagram showing the counter-streaming electron beam and plasma is depicted in Fig.~\ref{paper WI}.

The beam electron bends towards the $+x$-direction where the magnetic field is in the $-z$-direction due to the Lorentz force $q\Vec{u}_d\times \Vec{B}_z$, and bends towards the $-x$-direction where the magnetic field is in the $+z$-direction. The same applies to the background plasma electrons.

This bunches the electrons, generating high-density current filaments and current-depleted voids between them. The magnetic fields due to the beam current and plasma current add constructively. This process enhances the magnetic field: more filamentation leads to more field growth, which drives further filamentation. This positive feedback loop is the Weibel instability and causes the exponential growth of the magnetic field. Previous works have often kept ions as a non-evolving background during the saturation of the Weibel instability~\cite{cagas:2018}. Unlike the approach of Cagas \textit{et al.}~\cite{cagas:2018}, which treated ions as a non-evolving neutralizing background, the present work treats ions as a fully kinetic species and follows the system through the nonlinear ion-Weibel regime. In the case of cold counter-streaming plasma beams, the electric field creates potential wells comparable to the magnetic field. This results in saturation of the instability due to both electric and magnetic trapping. However, in hot counter-streaming plasma beams, the electric field is feeble compared to the magnetic field. Frederiksen~\cite{frederiksen2004magnetic} showed that this type of counter-streaming plasma is relevant for relative motions of sheets and filaments of galaxies and exists at large scales. Weibel-type instabilities are capable of creating the seed field for a large-scale magnetic field. According to Br\"{u}ggen~\textit{et al.}~\cite{bruggen2005simulations}, plasma turbulence excited during large-scale structure formation can amplify a seed field as strong as $B\ge10^{-10}$~G. The ion Weibel instability plays an important role in the merging of ion current channels. For hot beams, magnetic trapping~\cite{davidson1972nonlinear} is one of the principal candidates for the saturation of the Weibel instability. The saturation of the magnetic field in cold populations occurs from the establishment of potential wells that act in opposition to the filamentation force, thereby slowing down the expansion of the instability. The origin of these wells is attributed to a confluence of magnetic and electrostatic potentials~\cite{cagas2017nonlinear}. The overall magnetic energy increases as ion channels combine, and magnetic field patterns spread downstream. Electrons quickly reach thermal equilibrium, whereas ions maintain distinct bulk velocities in shielded ion channels and undergo slower thermalization~\cite{califano1998kinetic, califano1997spatial}. The work presented here shows the saturation of the Weibel instability in the presence of non-stationary ions through a detailed study of the conversion of initial kinetic energy into magnetic energy. Here we consider counter-streaming beams of both electrons and ions. The plasma beams are perturbed by a magnetic field perpendicular to the beams. The results from the two single-species cases and the four different temperature combinations for the multi-species case are presented. The Weibel instability induced by two symmetric counter-streaming ion beams has been discussed by some authors~\cite{ruyer:2015, jikei:2024, zhou:2023}. However, those studies considered the ion beam Weibel instability in weakly magnetized plasma with a finite background magnetic field, and discussed only the linear and weakly nonlinear phases of the instability. The present work, by contrast, treats fully unmagnetized plasma with evolving ion distributions and follows the system deep into the nonlinear saturation regime. We are not aware of a prior 1X2V continuum Vlasov-Maxwell study that simultaneously examines all four electron-ion temperature combinations with fully kinetic, evolving ions followed through the nonlinear saturation regime. This paper is organized as follows: Section~\ref{problem setup} sets up the problem using the Vlasov-Maxwell system of equations. Section~\ref{linear theory} derives the dispersion relation and solves it numerically and analytically. Section~\ref{simulation result} presents the simulation results obtained by directly solving the full Vlasov-Maxwell system using the continuum kinetic model under the \texttt{Gkeyll} framework~\cite{juno2018discontinuous,juno:2020,hakim:2020,francisquez:2024,mukherjee:2019,juno:2020a,juno:2021,francisquez:2020}. Section~\ref{observations} presents the observational context using Wind/SWE and MMS1 data. Section~\ref{Summary} summarizes our findings.

\section{Problem Setup}
\label{problem setup}

The Vlasov-Maxwell system of equations for electron and ion species is

\begin{equation}
\frac{\partial f_s}{\partial t} + \vec{v}\cdot\frac{\partial f_s}{\partial \vec{x}}
+ \frac{q_s}{m_s}\left(\vec{E}+\vec{v}\times\vec{B}\right)\cdot\frac{\partial f_s}{\partial \vec{v}}
= 0,
\label{VMeq}
\end{equation}

\begin{equation}
\frac{\partial \vec{B}}{\partial t} + \vec{\nabla}_{\vec{x}} \times \vec{E} = 0
\label{faraday paper}
\end{equation}

and

\begin{equation}
\varepsilon_0\mu_0\frac{\partial \vec{E}}{\partial t} - \vec{\nabla}_{\vec{x}} \times \vec{B}
= -\mu_0 \vec{J},
\label{faraday 2 paper}
\end{equation}

where $f_s$ is the distribution function for species $s$; $q_s$ and $m_s$ are the charge and mass of species $s$, respectively. $\Vec{E}$ and $\Vec{B}$ are the electric and magnetic fields, evolved using Maxwell's equations above. The operators $\vec{\nabla}_{\vec{x}}$ and $\vec{\nabla}_{\vec{v}}$ are the gradient operators in configuration and velocity space, respectively. We performed simulations of the system of equations~(\ref{VMeq}), (\ref{faraday paper}) and (\ref{faraday 2 paper}) under the \texttt{Gkeyll} framework with the initial distribution function for each plasma species taken as

\begin{equation}
\begin{split}
f_0(v_x, v_y) &= \sqrt{\frac{1}{2\pi v^2_{ths}}}\,e^{-\frac{v^2_x}{2v^2_{ths}}}
\times \frac{1}{2}\sqrt{\frac{1}{2\pi v_{ths}^2}}\\
&\quad \times \left(e^{-\frac{(v_y-u_d)^2}{2 v_{ths}^2}}
+e^{-\frac{(v_y+u_d)^2}{2 v_{ths}^2}}\right),
\end{split}
\label{maxwellian}
\end{equation}

where $v_{ths}=\sqrt{k_BT_s/m_s}$ is the thermal velocity of species $s$. Here we assume two counter-streaming plasma beams (with drift velocity $\Vec{u}_d$) in the $\pm y$-directions, with equal temperature $k_BT =\frac{1}{2}mv^2_{ths}$ in both directions. The \texttt{Gkeyll} framework employs a discontinuous Galerkin (DG) scheme with the serendipity basis set to discretize the phase-space advection and strong-stability-preserving Runge-Kutta time-steppers (SSP-RK) to discretize the time derivative~\cite{juno2018discontinuous,juno:2020,hakim:2020,francisquez:2024,mukherjee:2019,juno:2020a,juno:2021,francisquez:2020}. The DG scheme respects energy conservation well (derivation given in Appendix~\ref{dg scheme appendix}),

\begin{equation}
\frac{\partial}{\partial t}\left(\int\!\!\int \frac{1}{2}mv^2 f(x,v,t)\,dv\,dx\right)
+ \int\frac{\epsilon_0}{2}E^2\,dx + \int\frac{B^2}{2\mu_0}\,dx = 0.
\end{equation}

To simulate the full Vlasov-Maxwell equations, we used a continuum kinetic method that directly evolves the particle distribution function in phase space. Kinetic simulations evolve the distribution functions of particles in velocity space, allowing for a detailed description of particle kinetics. As continuum kinetic simulations are noise-free, a perturbation initialized purely in the perpendicular direction remains perpendicular. We consider an initially unmagnetized plasma system made of counter-streaming populations of charged particles with uniform density and velocity profiles. The charged particle beams flow with drift velocity $\pm u_d=0.1c$ in the $y$-direction. The simulations use the physical ion-to-electron mass ratio $m_i/m_e = 1836$; the qualitative ordering of thermalization timescales between species is not sensitive to this choice at the thermal-to-drift velocity ratios studied here. First, we consider counter-streaming electron beams with ions forming a stationary, non-evolving neutralizing background, i.e., $u_{di}=0$ on the timescales of interest. Next, we consider counter-streaming electron-ion (proton) beams in which ions evolve on plasma frequency timescales with ion distribution function $f_{i}$ evolving with the perturbation. The density and velocity of the plasma species are taken to preserve quasi-neutrality. The initial uniform but unstable equilibrium is disturbed with the magnetic perturbation ($B_z$) introduced in the $z$-direction as

\begin{equation}
B_{z}(x) = B_{z,0}\sin(k_{0}x),
\label{perturbation}
\end{equation}

where $k_0$ is the wavenumber of the initial perturbation and $B_{z,0}=10^{-3}$ in normalized units. Periodic boundary conditions are used with the system length spanning from $-2\pi/k_0$ to $2\pi/k_0$. The chosen domain length of $4\pi/k_0$ accommodates exactly two wavelengths of the seed mode; sub-harmonic modes at $k_0/2$, which are relevant to filament merging on large scales, cannot fit within this domain and are therefore excluded from the analysis. The simulation was carried out with 64 spatial grid points and $256\times256$ velocity-space grid cells. The simulation results are presented in Section~\ref{simulation result}.

Throughout this paper, velocities are normalized to $c$, time to $\omega_{pe}^{-1}$, lengths to $c/\omega_{pe}$, and fields to $m_e\omega_{pe}c/e$.

\section{Linear Theory}
\label{linear theory}

Before presenting simulation results, we study the linear instability analysis of the system of equations~(\ref{VMeq}), (\ref{faraday paper}) and (\ref{faraday 2 paper}). The velocity distribution functions of both species are found by linearizing the Vlasov equation~(\ref{VMeq}). The linearization process is given in Appendix~\ref{app A}.

\begin{equation}
\begin{split}
f_{s,1} &= \frac{-iq_s}{m_{s}(\omega-v_{x}k_x)}
\left[(E_{x,1}+v_{y}B_{z,1})\frac{\partial f_{s,0}}{\partial v_x}\right.\\
&\quad\left. + (E_{y,1}- v_{x}B_{z,1})\frac{\partial f_{s,0}}{\partial v_y}\right]
\end{split}
\label{linearized_max}
\end{equation}

The linearized Amp\`{e}re's law is

\begin{equation}
\begin{split}
ik_{x}B_{z,1} &= -\frac{i\omega}{c^2}E_{y,1}\\
&\quad + \mu_{0}\left[q_{e}\int v_{y}f_{e,1}\,dv
+ q_{i}\int v_{y}f_{i,1}\,dv\right].
\end{split}
\label{linearized_amperes_law}
\end{equation}

Combining equations~(\ref{linearized_max}) and (\ref{linearized_amperes_law}) yields the following dispersion relation (derived in Appendix~\ref{app B}):

\begin{equation}
\begin{split}
&1-\frac{\omega_{pe}^2}{k^2c^2}\left[\frac{u^2_{de}}{v^2_{the}}
+\zeta_e Z(\zeta_e)\left(1 + \frac{u^2_{de}}{v^2_{the}}\right)\right]\\
&-\frac{\omega^2_{pi}}{k^2c^2}\left[\frac{u^2_{di}}{v^2_{thi}}
+\zeta_i Z(\zeta_i)\left(1 + \frac{u^2_{di}}{v^2_{thi}}\right)\right]
- \frac{\omega^2}{k^2c^2} = 0,
\end{split}
\label{disp}
\end{equation}

where $\omega_{pe}$ and $\omega_{pi}$ are the electron and ion plasma frequencies, respectively, and $k$ is the instability wavenumber. $Z(\zeta_s)$ is the plasma dispersion function defined as

\begin{equation}
Z(\zeta_s)=\pi^{{-1}/{2}}\int_{-\infty}^{\infty}\frac{\exp(-t^2)}{t-\zeta_s}\,dt,
\label{plasma dispersion function}
\end{equation}

where $\zeta_s=\frac{\omega}{\sqrt{2}v_{ths}k}$ and $\omega =\omega_r+i\gamma$ with $\gamma$ as the instability growth rate. We numerically solved the dispersion relation~(\ref{disp}) in two ways: (i) asymptotic expansion for large $\zeta$ and power series expansion for small $\zeta$ in the plasma dispersion function~(\ref{plasma dispersion function}), and (ii) using the plasma dispersion function module in Python to solve equation~(\ref{plasma dispersion function}) directly. For clarity in the result analysis, we refer to the first approach as Theory~I and the second as Theory~II. In the following section we present results for both single-species and two-species plasma under different temperature conditions.

\begin{table}[t]
\caption{\label{tab:table1} Simulation parameters for single-species case}
\begin{ruledtabular}
\begin{tabular}{ccc}
 & HE\footnote{HE = Hot electron} & CE\footnote{CE = Cold electron} \\
\hline
$v_{the}/u_d$ & $1.0$ & $0.1$ \\
$u_d/c$       & $0.1$ & $0.1$ \\
\end{tabular}
\end{ruledtabular}
\label{simulaton overview single species}
\end{table}

\begin{table}[t]
\caption{\label{tab:table2} Simulation parameters for two-species case\footnote{Note that the hot electron parameter in the single-species case ($v_{the}/u_d = 1.0$, Table~\ref{tab:table1}) differs from the multi-species hot cases ($v_{the}/u_d = 0.5$, this table); both regimes satisfy $v_{the} \geq u_d$ and exhibit magnetically dominated saturation.}}
\begin{ruledtabular}
\begin{tabular}{ccccc}
& CECI\footnote{CECI = Cold electron and cold ion}
& CEHI\footnote{CEHI = Cold electron and hot ion}
& HECI\footnote{HECI = Hot electron and cold ion}
& HEHI\footnote{HEHI = Hot electron and hot ion} \\
\hline
$v_{the}/u_d$ & $0.1$ & $0.1$ & $0.5$ & $0.5$ \\
$v_{thi}/u_d$ & $0.1$ & $0.5$ & $0.1$ & $0.5$ \\
$u_d/c$       & $0.1$ & $0.1$ & $0.1$ & $0.1$ \\
\end{tabular}
\end{ruledtabular}
\label{simulaton overview}
\end{table}

\begin{figure}[h!]
\centering
\includegraphics[width=1\linewidth]{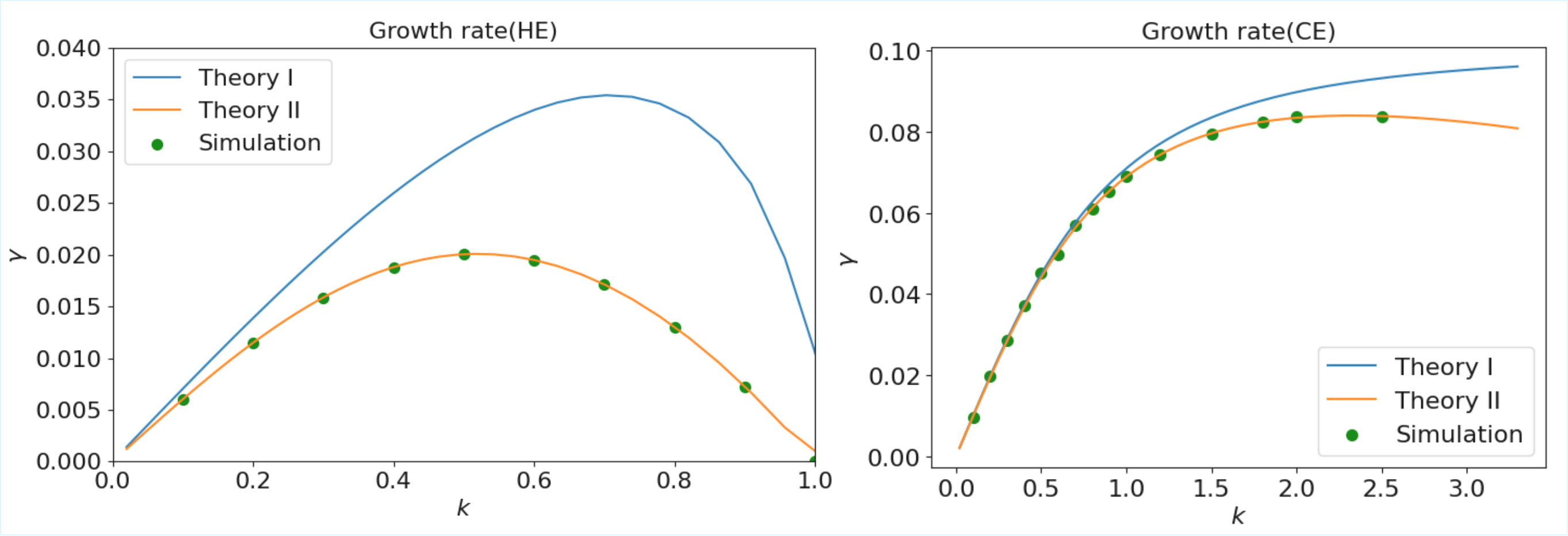}
\caption{Comparison of growth rates from simulation and linear theory for the single-species case. \textit{Left panel}: hot single-species electron beams (HE), showing good agreement between Theory~I, Theory~II, and simulation. \textit{Right panel}: cold single-species electron beams (CE), showing even closer agreement, consistent with the cold plasma approximation being more accurate in this regime. In both cases, ions form a stationary background.}
\label{singlecoldanalytic}
\end{figure}

\section{Simulation Results}
\label{simulation result}

To study the detailed behavior of nonlinear saturation of the Weibel instability in counter-streaming electron-ion plasma, we divided our analysis into three approaches as described in the previous section: (i) the dispersion relation using asymptotic and power series expansions of the plasma dispersion function (Theory~I), (ii) the dispersion relation using the plasma dispersion function module in Python (Theory~II), and (iii) the Vlasov-Maxwell system of equations using the \texttt{Gkeyll} framework (Simulation). The analysis was carried out in one spatial dimension and two velocity dimensions (1X2V) for low- and high-temperature beams in single-species and two-species plasma (electrons and singly charged positive ions). This section is organized into two subsections: the single-species case (Section~\ref{single species section}) and the two-species case (Section~\ref{multi species section}). Each subsection is further divided according to whether the counter-streaming charged particles are cold or hot.

\subsection{Single-species case}
\label{single species section}

In the single-species case, we consider two counter-streaming electron beams with a stationary ion background, perturbed by a magnetic field perpendicular to the beam direction. The ion drift velocity is taken to be zero ($u_{di}=0$). Two sub-cases are considered: hot counter-streaming electron beams with $v_{the}/u_{de}=1.0$, and cold counter-streaming electron beams with $v_{the}/u_{de}=0.1$.

\subsubsection{Single-species hot electron beams (HE)}
\label{singlehot}

\begin{figure}[h!]
\centering
\includegraphics[width=1\linewidth]{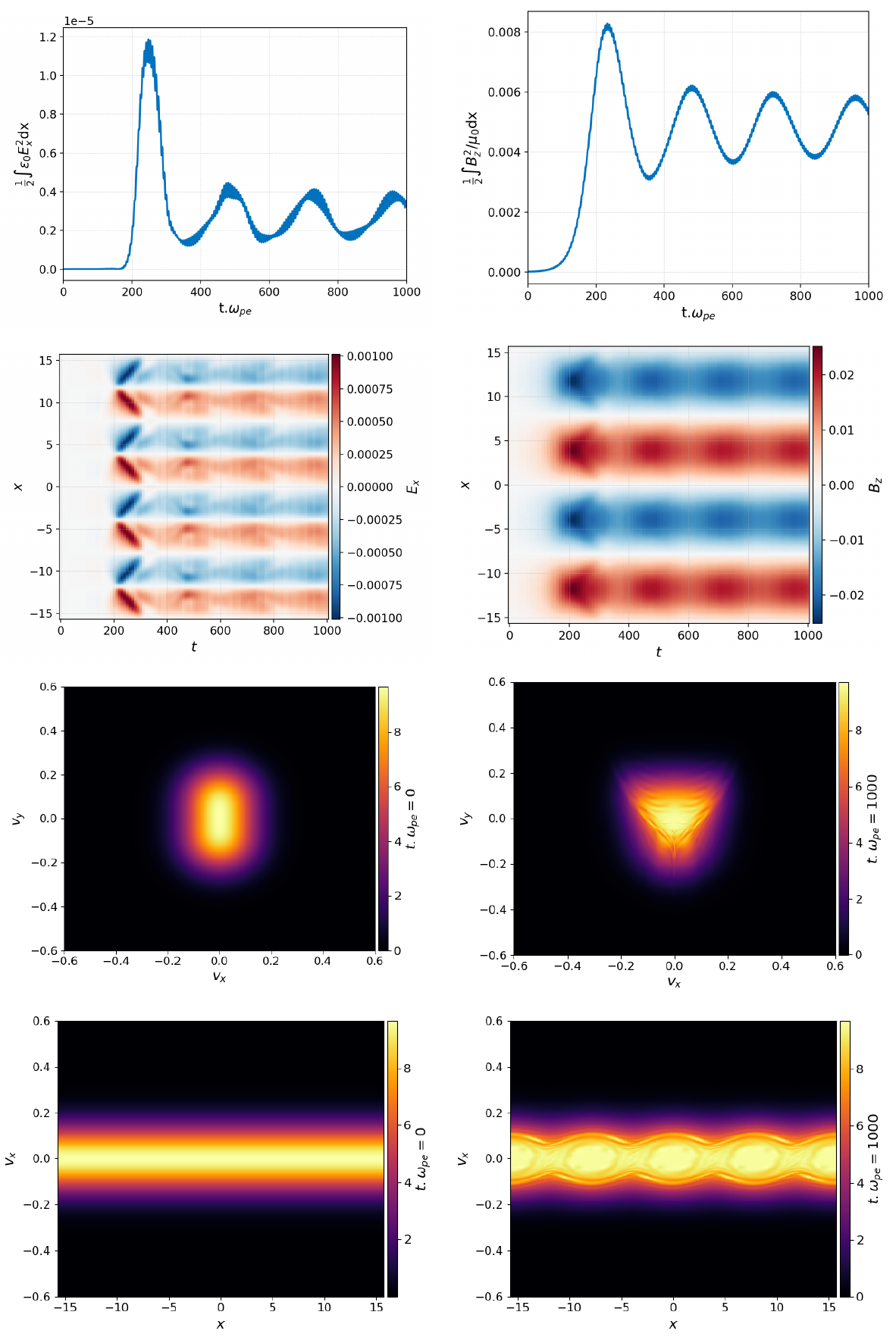}
\caption{Single-species high-temperature plasma beams (HE case). \textit{Top panel}: time evolution of electric field energy (left)
and magnetic field energy (right). \textit{Second panel}: spatio-temporal evolution of $E_x$ (left) and $B_z$ (right).
\textit{Third panel}: velocity distribution in $V_x$--$V_y$ space at $t\omega_{pe}=0$ (left) and $t\omega_{pe}=1000$ (right), showing
beam merging into a single broadened population. \textit{Bottom panel}: phase-space distribution in $x$--$V_x$ at $t\omega_{pe}=0$
(left) and $t\omega_{pe}=1000$ (right), showing particle trapping and phase mixing.}
\label{hot single}
\end{figure}

For the single-species case, the kinetic dispersion relation~(\ref{disp}) reduces to

\begin{equation}
\frac{\omega_{pe}^2}{c^2k^2}\left[\zeta Z(\zeta) \left(1 + \frac{u^2_d}{v^2_{the}}\right)
+ \frac{u^2_d}{v^2_{the}} \right]  + \frac{2v^2_{the}}{c^2} \zeta^2=1.
\label{single species dispersion relation}
\end{equation}

Assuming electrons are hot enough that $\zeta=\frac{\omega}{\sqrt{2}v_{the}k}\ll 1$, we use the power series expansion for small argument in $Z(\zeta)$ to obtain the approximate dispersion relation from equation~(\ref{single species dispersion relation}) as

\begin{equation}
\omega^2 \left[1 - \frac{\omega^2_{pe}}{v^2_{the}k^2}
\left(1 + \frac{u^2_{de}}{v^2_{the}} \right) \right]
+ \left[\omega^2_{pe} \frac{u^2_{de}}{v^2_{the}} - c^2k^2 \right] = 0.
\end{equation}

Applying the purely imaginary character of the Weibel mode ($\omega=i\gamma$ with $\gamma>0$), the growth rate $\gamma$ is

\begin{equation}
\gamma=\omega_{pe}\sqrt{
\frac{c^2k^2-\dfrac{\omega_{pe}^2u_d^2}{v_{the}^2}}
{\dfrac{\omega_{pe}^2}{v_{the}^2 k^2}
\!\left( 1 + \dfrac{u_d^2}{v_{the}^2} \right) - 1}}\,.
\label{analyticalgrowthratehot}
\end{equation}

The approximate growth rate (equation~\ref{analyticalgrowthratehot}) is plotted for different wavenumbers $k$ with the parameters in Table~\ref{simulaton overview single species} and labeled `Theory~I' in Fig.~\ref{singlecoldanalytic} (left panel). We also numerically solved equation~(\ref{single species dispersion relation}) using the plasma dispersion function $Z(\zeta)$ module in Python using SciPy's \texttt{fsolve}, which iteratively refines the roots for each wavenumber. These results are labeled `Theory~II'. The kinetic simulation results are labeled `Simulation'. From Fig.~\ref{singlecoldanalytic}, the cold case (CE) has a higher peak growth rate than the hot case (HE), while the hot case grows more slowly in time due to the broader velocity distribution suppressing filamentation. Theory~I, despite using approximations, also captures the overall shape and the location and amplitude of the dominant mode well. In the presence of the magnetic field perturbation $B_z$, each electron beam is deflected in transverse opposite directions ($\pm u_x$) by the force $qu_dB_z$. This generates transverse filamentation currents, driving the growth of the magnetic field (Fig.~\ref{hot single}). Both magnetic and electric field energies grow, converted from the free kinetic energy of the electrons. Due to the 1X2V geometry, the magnetic field grows only in the $z$-direction, while the electric field grows in $x$ and $y$: $\vec{E}(x) = E_x \hat{x} + E_y \hat{y}$. However, the $E_x$ component dominates over $E_y$~\cite{cagas2017nonlinear}. The nonlinear saturation of the Weibel instability for single-species plasma in this setup was studied by Cagas \textit{et al.}~\cite{cagas2017nonlinear}, whose results we reproduce here as a reference baseline for the multi-species comparisons that follow. The magnetic field grows exponentially from the initial perturbation until it reaches nonlinear saturation (Fig.~\ref{hot single}, first row). We observe oscillatory behavior of the field energy in the nonlinear phase: during linear growth, the filamentation force generates transverse velocity $u_x$; once saturation is reached, $u_x$ vanishes and the electric field decays, but the filamentation force reintroduces $u_x$, leading to a second saturation. This produces the periodic nonlinear oscillations seen in the figure. The growth of the electric field energy is much smaller than the magnetic field energy, indicating that magnetic trapping is the dominant saturation mechanism for hot electrons, consistent with Cagas \textit{et al.}~\cite{cagas2017nonlinear}. The second row of Fig.~\ref{hot single} shows the evolution of phase space at $t\omega_{pe}=0$ and $t\omega_{pe}=100$. Initially the two beams form separate structures in velocity space; by saturation the beams merge into a single anisotropic population with broader distribution in $v_x$. The last row at $t\omega_{pe}=300$ shows particle trapping and phase mixing, the latter causing significant broadening of the distribution function and plasma heating during the instability evolution.

\subsubsection{Single-species cold electron beams (CE)}
\label{singlecold}

\begin{figure}[h!]
\centering
\includegraphics[width=1\linewidth]{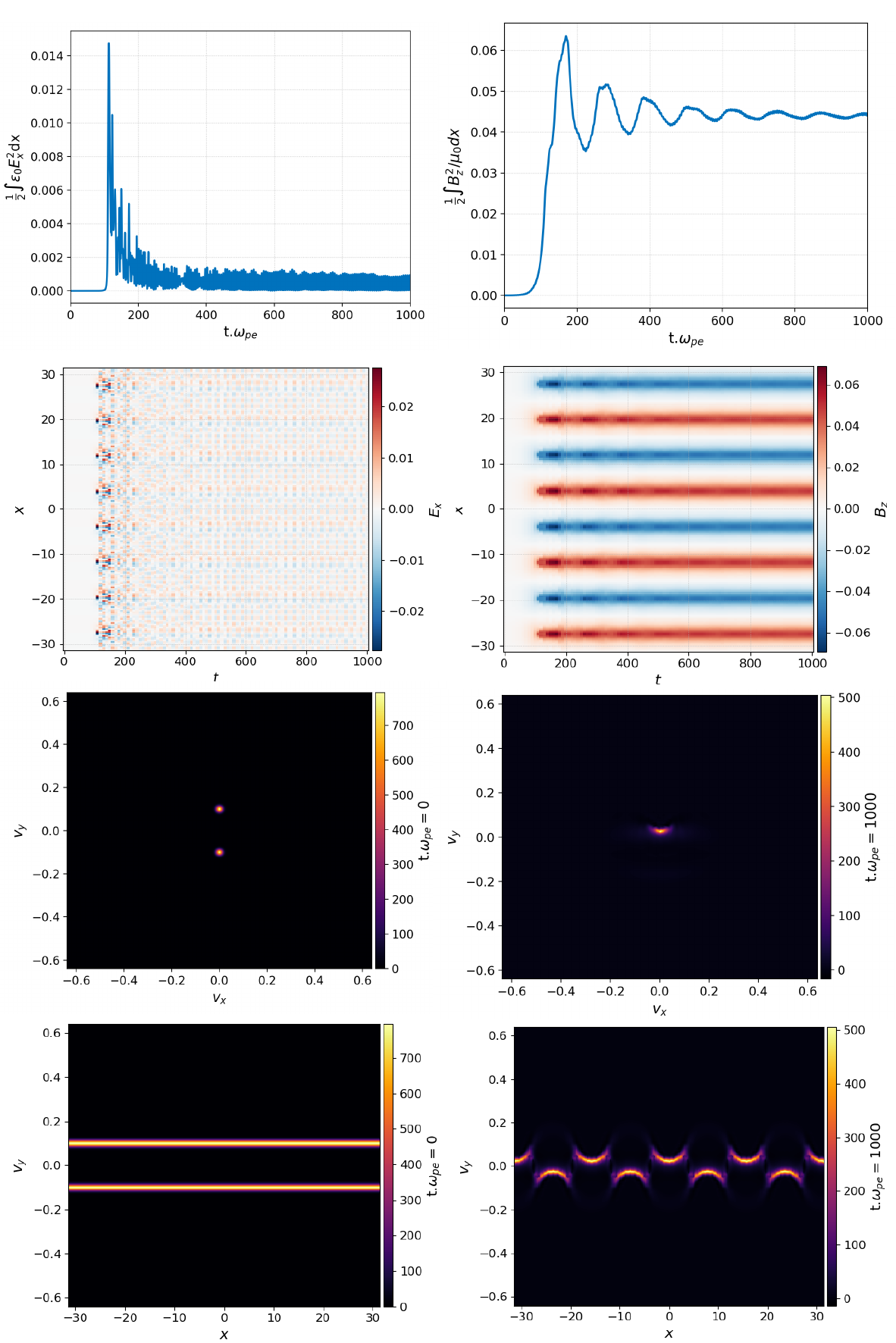}
\caption{Single-species cold-temperature plasma beams (CE case). \textit{Top panel}: time evolution of electric field energy (left) and magnetic field energy (right). \textit{Second panel}: spatio-temporal evolution of $E_x$ (left) and $B_z$ (right). \textit{Third panel}: velocity distribution in $V_x$--$V_y$ space at $t\omega_{pe}=0$ (left) and $t\omega_{pe}=1000$ (right). \textit{Bottom panel}: phase-space distribution in $x$--$V_x$ at $t\omega_{pe}=0$ (left) and $t\omega_{pe}=1000$ (right), showing particle trapping and beam filamentation.}
\label{cold single}
\end{figure}

For initially unmagnetized cold plasma with a stationary ion background, the parameter $\zeta\gg 1$. Using the asymptotic expansion of $Z(\zeta)$ for $\zeta\gg1$, the cold Weibel dispersion relation is obtained from equation~(\ref{disp}) as

\begin{equation}
\omega^4-\omega^2(\omega^2_{pe}+c^2k_x^2)-\omega^2_{pe}u_d^2k_x^2=0.
\label{cold single disp rel}
\end{equation}

Solving for $\omega$, the four roots are

\begin{equation}
\omega =\pm\sqrt{ \frac{1}{2} \left[(\omega^2_{pe} + c^2k^2)
\pm \sqrt{(\omega^2_{pe} + c^2k^2)^2 + 4\omega^2_{pe}u^2_dk^2} \right]}.
\label{four root}
\end{equation}

Of the four roots, two are light modes $(+,+)$ and $(-,+)$, and two are Weibel modes $(+,-)$ and $(-,-)$. Taking $(+,-)$ for the maximum growth rate $\gamma$,

\begin{equation}
\gamma =\sqrt{ \frac{1}{2} \left[ \sqrt{(\omega^2_{pe} + c^2k^2)^2 + 4\omega^2_{pe}u^2_dk^2}
-(\omega^2_{pe} + c^2k^2) \right]}.
\label{gamma for single cold}
\end{equation}

The growth rate in equation~(\ref{gamma for single cold}) is plotted as `Theory~I' in Fig.~\ref{singlecoldanalytic} (right panel). Theory~II and the simulation are calculated with the same parameters. The simulation and theoretical results agree well at low wavenumbers before nonlinear saturation sets in. The overall agreement between theory and simulation is better in the cold plasma case compared to the hot case. Figure~\ref{cold single} (first row) shows the evolution of electric and magnetic field energies. Unlike in the hot case (Fig.~\ref{hot single}), the oscillatory behavior is not pronounced in the cold case. Notably, both the electric and magnetic field energies grow to comparable magnitudes. This indicates that an electrostatic potential plays an important role alongside magnetic trapping: the electrostatic field creates potential wells that accumulate electrons and contribute to saturation of the instability. The second panel shows the evolution of the phase-space distribution from the initial condition through the nonlinear regime.

\subsection{Multi-species case}
\label{multi species section}

Here we consider an initially unmagnetized plasma with counter-streaming electron and proton beams. We study the generation of magnetic fields from zero initial field, initiated by the transverse magnetic perturbation given in equation~(\ref{perturbation}). Since both electrons and ions move along the $y$-direction and produce equal particle fluxes in opposite directions, the net current is zero. The initial magnetic perturbation produces a Lorentz force that deflects the electron trajectories first (Fig.~\ref{paper WI}). As a result, electrons moving upward concentrate on one side and those moving downward on the other, forming a current sheath. This current sheath amplifies the magnetic field perturbation. The instability in this case is driven by particle anisotropy and vanishes as the system tends toward the isotropic state. As the magnetic field grows, the Lorentz deflections scatter the particle pitch angles, driving the system toward isotropy. If the initial condition has strong anisotropy, where the thermal spread is much smaller than the bulk velocity, the particles will eventually isotropize such that the thermal energy equals the initial directed kinetic energy. This final state marks the saturation of the instability.

\begin{figure}[h!]
\centering
\includegraphics[width=\linewidth]{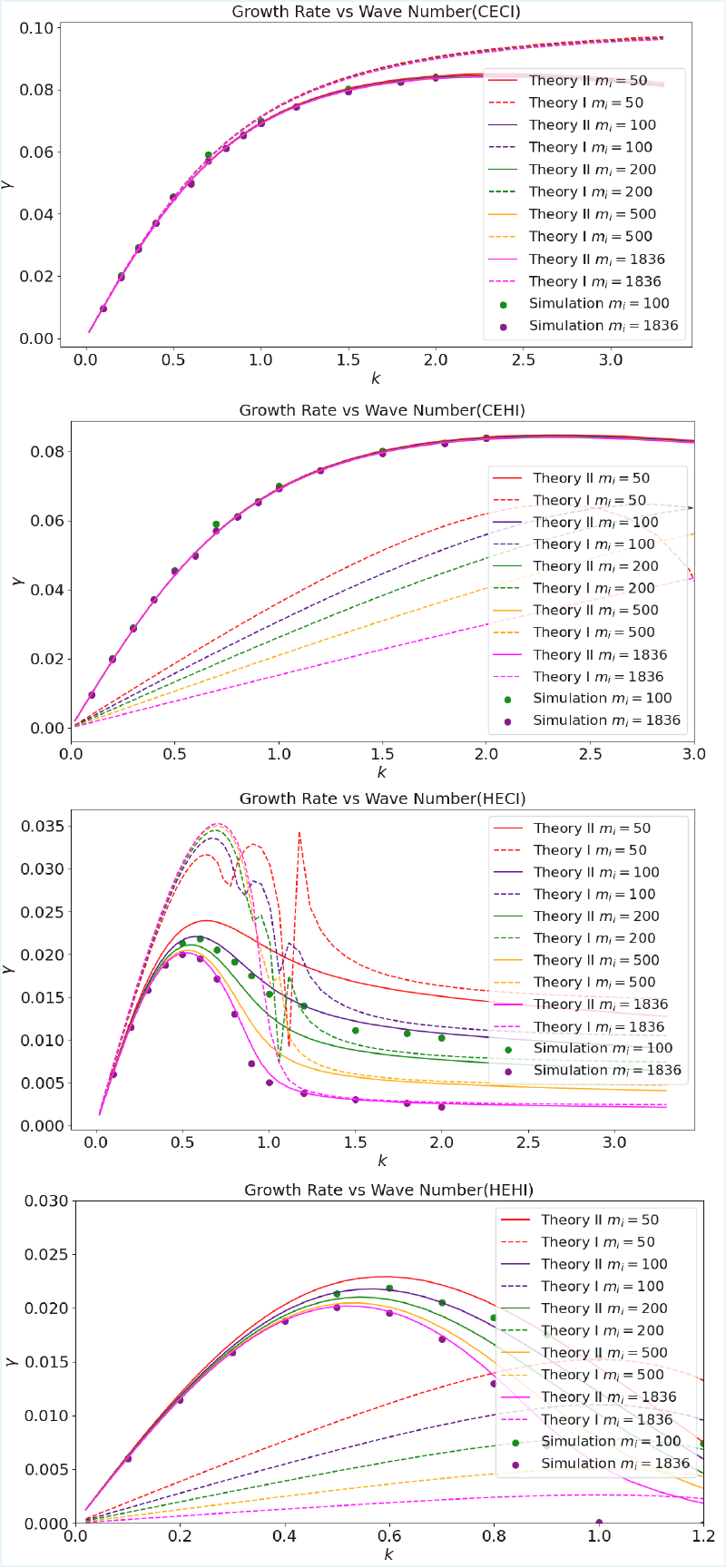}
\caption{Growth rates from simulation and linear theory for the multi-species cases. From top to bottom: CECI (cold electron, cold ion), CEHI (cold electron, hot ion), HECI (hot electron, cold ion), and HEHI (hot electron, hot ion). The growth rate patterns for CECI and CEHI resemble each other, as do those for HECI and HEHI, consistent with the linear growth rate being governed primarily by the electron temperature.}
\label{multiple analytic growth rate}
\end{figure}

\subsubsection{Growth rate}

We divide our study into four categories according to the electron and proton temperatures as given in Table~\ref{simulaton overview}: cold electron cold ion (CECI), cold electron hot ion (CEHI), hot electron cold ion (HECI), and hot electron hot ion (HEHI). Considering the asymptotic expansion (for cold species) and the small-argument power series (for hot species), we have $Z(\zeta_s)\sim -\frac{1}{\zeta_s}-\frac{1}{2\zeta^3_s}$ and $Z(\zeta_s)\sim -2\zeta_s + \frac{4\zeta^3_s}{3}$ respectively, where $\zeta_s = \frac{\omega}{\sqrt{2}v_{ths}k}$. Using these approximations, equation~(\ref{disp}) becomes for CECI, HECI, HEHI, and CEHI, respectively,

\begin{equation}
\begin{split}
&\omega^4-\omega^2(k^2c^2 + \omega^2_{pe}+\omega^2_{pi})
- \omega^2_{pe}u^2_{de}k^2-\omega^2_{pi}u^2_{di}k^2 \\
&-\omega^2_{pe}k^2v^2_{the}-\omega^2_{pi}k^2v^2_{thi} = 0,
\end{split}
\label{ceci dispersion relation}
\end{equation}

\begin{equation}
\begin{split}
& \omega^4\!\left(1-\frac{\omega^2_{pe}}{v_{the}^2k^2}
-\frac{\omega^2_{pe}u_{de}^2}{v_{the}^4k^2}\right)
-\omega^2\!\left(k^2c^2-\frac{\omega^2_{pe}u_{de}^2}{v_{the}^2}+\omega^2_{pi}\right)\\
& -\left(\omega^2_{pi}v^2_{thi}k^2+\omega^2_{pi}u^2_{di}k^2\right)=0,
\end{split}
\label{heci dispersion relation}
\end{equation}

\begin{equation}
\begin{split}
& \omega^4\!\left(\frac{\omega_{pe}^2}{3v^4_{the}k^4}
+\frac{\omega_{pe}^2 u_{de}^2}{3v^6_{the}k^4}
+\frac{\omega_{pi}^2}{3v^4_{thi}k^4}
+\frac{\omega_{pi}^2 u_{di}^2}{3v^6_{thi}k^4}\right)\\
& -\omega^2\!\left(-1+\frac{\omega_{pe}^2}{v^2_{the}k^2}
+\frac{\omega_{pe}^2u_{de}^2}{v^4_{the}k^2}
+\frac{\omega_{pi}^2}{v^2_{thi}k^2}
+\frac{\omega_{pi}^2u_{di}^2}{v^4_{thi}k^2}\right)\\
& +\left(-k^2c^2+\frac{\omega_{pe}^2u_{de}^2}{v^2_{the}}
+\frac{\omega_{pi}^2u_{di}^2}{v^2_{thi}}\right)=0
\end{split}
\label{hehi disp rel}
\end{equation}

and

\begin{equation}
\begin{split}
& \omega^4\!\left(1-\frac{\omega^2_{pi}}{v_{thi}^2k^2}
-\frac{\omega^2_{pi}u_{di}^2}{v_{thi}^4k^2}\right)
-\omega^2\!\left(k^2c^2-\frac{\omega^2_{pi}u_{di}^2}{v_{thi}^2}+\omega^2_{pe}\right)\\
& -\left(\omega^2_{pe}v^2_{the}k^2+\omega^2_{pe}u^2_{de}k^2\right)=0.
\end{split}
\label{cehi disp rel}
\end{equation}

Considering the purely imaginary character of the Weibel modes ($\omega=i\gamma$ with $\gamma>0$), the respective growth rates ($\gamma$) are derived in Appendices~\ref{app C}--\ref{app F}. From Fig.~\ref{multiple analytic growth rate}, the growth rate patterns for CECI and CEHI resemble each other, as do those for HECI and HEHI. This is consistent with our expectation that the linear growth rate is not strongly affected by the temperature or dynamics of the ions; it is the electron temperature that primarily governs the linear phase. The following subsections examine the nonlinear behavior of each case.

\subsubsection{Cold electron, cold ion (CECI)}
\label{ceci section}

\begin{figure}[h!]
\centering
\includegraphics[width=1\linewidth]{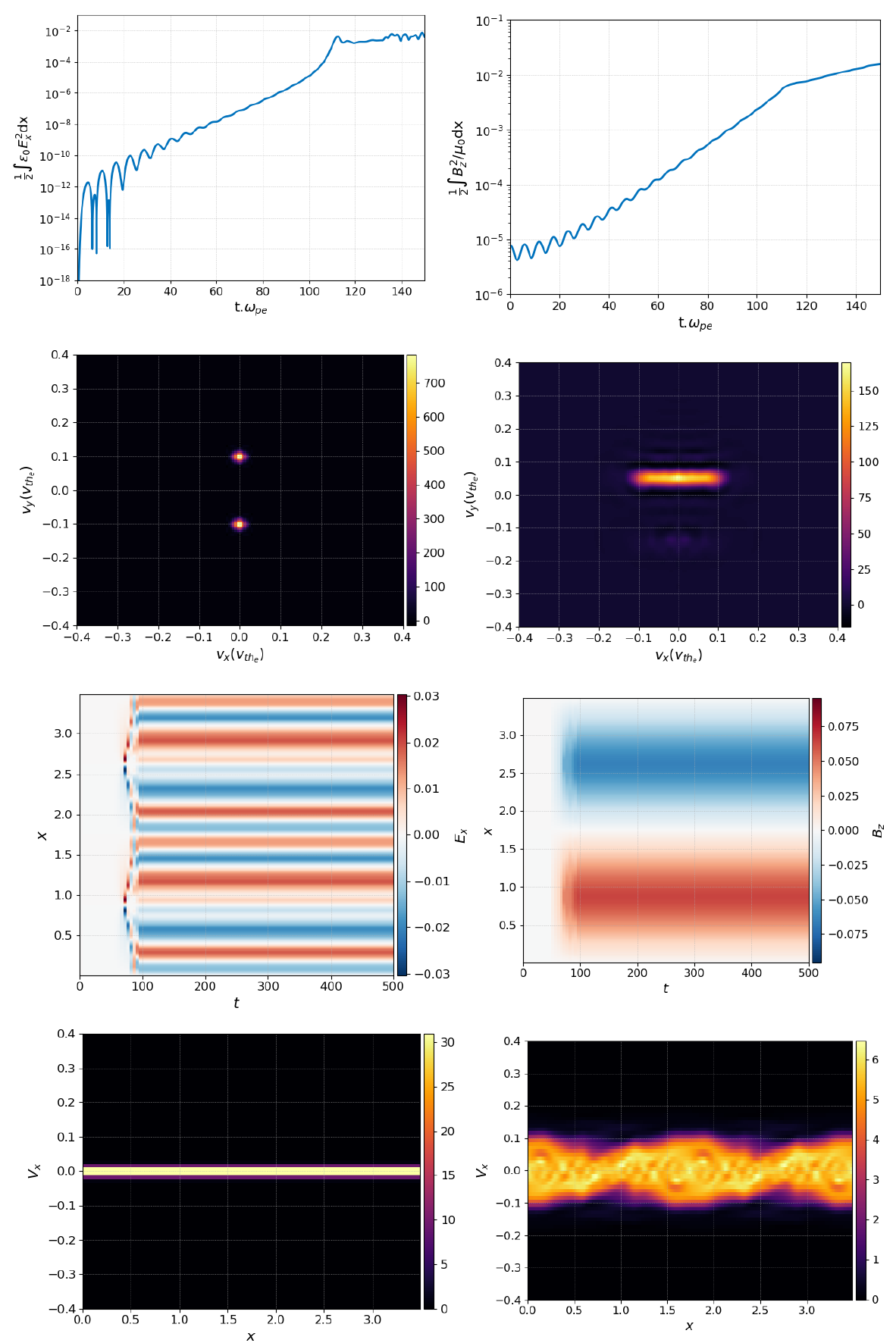}
\caption{Multi-species cold electron, cold ion (CECI) case. \textit{Top panel}: time evolution of electric field energy (left) and magnetic field energy (right). Both energies saturate at comparable levels after $t\omega_{pe}\simeq115$, consistent with the role of electrostatic potential wells alongside magnetic trapping. \textit{Second panel}: velocity distribution function in $V_x$--$V_y$ space at $t\omega_{pe}=0$ (left) and $t\omega_{pe}=100$ (right), showing beam merging. \textit{Third panel}: spatio-temporal evolution of electric field $E_x$ (left) and magnetic field $B_z$ (right). \textit{Bottom panel}: phase-space distribution in $x$--$V_x$ at $t\omega_{pe}=0$ (left) and $t\omega_{pe}=100$ (right), showing particle trapping and broadening of the distribution function due to phase mixing.}
\label{ceci}
\end{figure}

The simulation results for CECI are illustrated in Fig.~\ref{ceci}. From the top panel, the electric field grows with some noise until $t\omega_{pe}\simeq50$, after which the noise subsides; the electric field continues growing until $t\omega_{pe}\simeq115$ and then saturates. Although the electric and magnetic field energies saturate at nearly the same levels and timescales, the electric field energy is initially much smaller than the magnetic field energy; for example, at $t\omega_{pe}\simeq60$ the electric field energy is around $10^{-8}$ while the magnetic field energy is around $10^{-4}$. In the single-species cold case, the saturation level was already reached before $t\omega_{pe}\simeq60$ (Fig.~\ref{cold single}). Comparing Figs.~\ref{cold single} and \ref{ceci}, the similar growth patterns indicate that the cold electron dynamics dominate the early nonlinear evolution. The incoming electrons, being lighter than ions, are immediately deflected by fluctuations in the electric field, leading to the onset of the two-stream instability, which operates concurrently with the Weibel (filamentation) instability in the cold beam regime. Following Amp\`{e}re's law, cylindrical magnetic fields surround the current channels and promote mutual attraction between them. These channels merge to form larger ones, producing transverse magnetic fields and increasing the magnetic field energy. This progression continues until the magnetic field becomes sufficiently strong to divert the heavier ions into the magnetic voids between electron channels. The ions then undergo the same growth mechanism as the electrons, but at a much slower rate due to their heavier mass. As the ion channels strengthen, they begin to encounter shielding effects from electrons heated by the growing electromagnetic structures. The two electron beams, initially separated in velocity space, merge into a single population as seen in the second panel of Fig.~\ref{ceci}. The third panel shows that there is no significant growth of magnetic or electric fields until $t\omega_{pe}\simeq100$, consistent with the saturation timescale shown in the top panel. The bottom panel shows the distribution function integrated over $V_y$, projected onto $x$--$V_x$. It shows evidence of particle trapping and broadening of the distribution function along $V_x$, reflecting overall heating of the electron population due to phase mixing.

\subsubsection{Hot electron, cold ion (HECI)}
\label{heci section}

\begin{figure}[h!]
\centering
\includegraphics[width=1\linewidth]{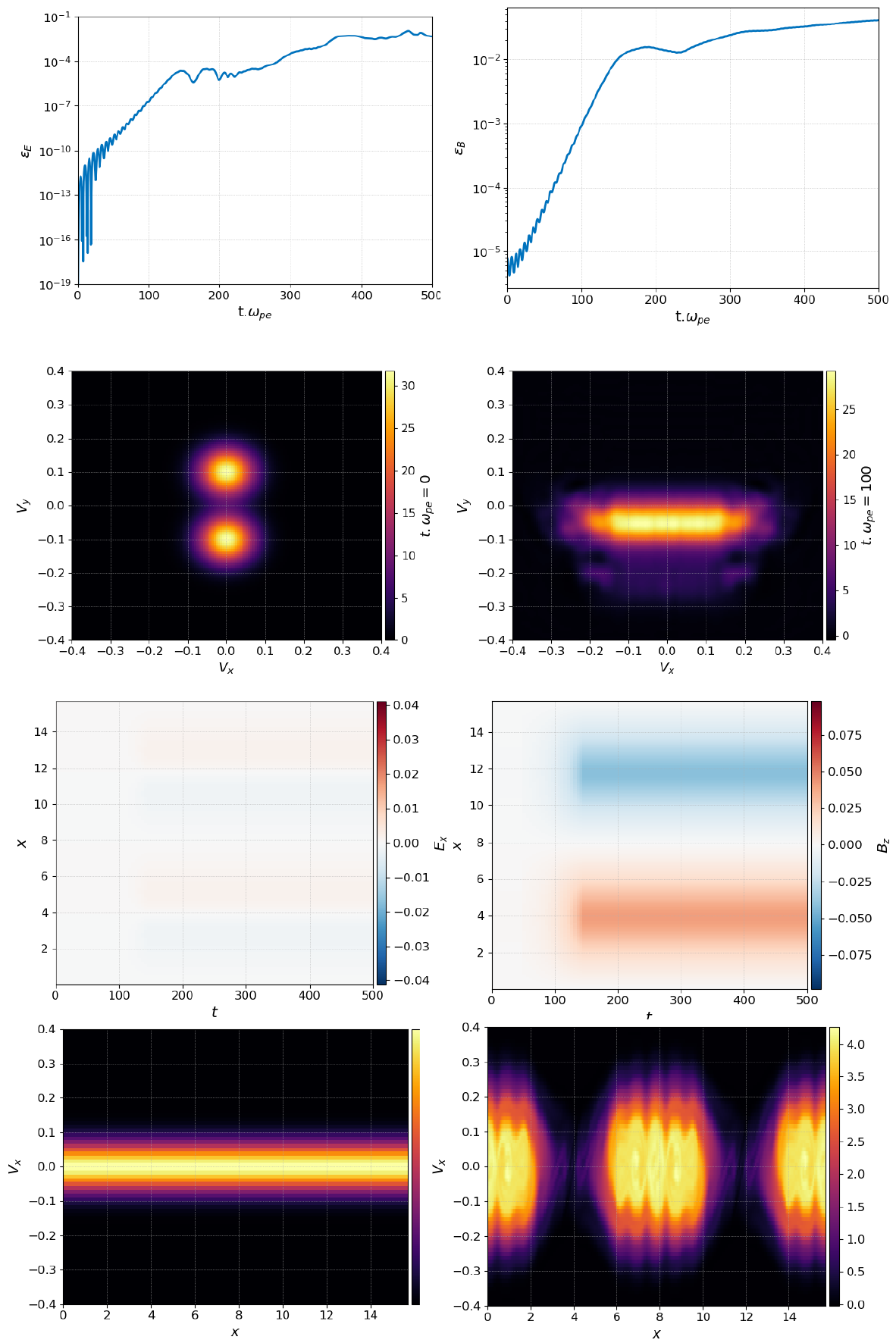}
\caption{Multi-species hot electron, cold ion (HECI) case. \textit{Top panel}: time evolution of electric field energy (left) and magnetic field energy (right). The magnetic field energy is substantially larger than the electric field energy throughout, consistent with magnetic trapping as the dominant saturation mechanism. \textit{Second panel}: velocity distribution function in $V_x$--$V_y$ space at $t\omega_{pe}=0$ (left) and $t\omega_{pe}=100$ (right). \textit{Third panel}: spatio-temporal evolution of electric field $E_x$ (left) and magnetic field $B_z$ (right), showing the spatial coherence of filamentation structures and their slow downstream spreading at late times. \textit{Bottom panel}: phase-space distribution in $x$--$V_x$ at $t\omega_{pe}=0$ (left) and $t\omega_{pe}=100$ (right).}
\label{heci}
\end{figure}

Here we consider hot electrons and cold ions. Because of their lower mass and higher thermal energy, electrons react to field perturbations much faster than the ions, which is why many previous studies ignored the ion dynamics altogether~\cite{cagas:2018, cagas2017nonlinear, frederiksen2004magnetic}. The present work, by treating ions as a fully kinetic species, shows the distinct late-time ion channel dynamics not apparent in those earlier models. From the simulation, the electric field grows with some noise until $t\omega_{pe}=70$, after which the noise disappears but the field continues growing until $t\omega_{pe}=147$, then shows a sharp dip, and finally saturates after $t\omega_{pe}=174$. The magnetic field energy shows a similar noisy initial growth until $t\omega_{pe}=67$ and finally grows to $t\omega_{pe}=158$, after which it saturates with very slow residual growth. The magnetic field energy is substantially larger than the electric field energy throughout (Fig.~\ref{heci}, third panel). The beam distribution quickly relaxes toward a plateau. The primary instability in the longitudinal direction is the two-stream instability, which produces an anisotropic bi-Maxwellian distribution function. Concurrently, the Weibel instability manifests as a secular instability within this bi-Maxwellian framework, generating magnetic fields through purely amplifying modes. When the spatial scale of the generated fields approaches the electron gyroradius, the magnetic fields stabilize as electrons are magnetically trapped within the wave potential. The second panel of Fig.~\ref{heci} shows that the electron populations merge into a single population over time. The ions follow the same mechanism but much more slowly due to their mass ratio and cold temperature. The magnetic energy density continues growing slowly after $t\omega_{pe}=158$ throughout the simulation. The third panel shows the spatio-temporal evolution of $E_x$ and $B_z$: at late times the magnetic filament structures spread along the spatial direction, consistent with the downstream field spreading described in the abstract. The ion-electron asymmetry in thermalization timescales plays an important part in sustaining the current channels. Over time, incoming electrons are scattered and eventually thermalized. The magnetic field makes the velocity distribution more isotropic, while the electric field arising from charge separation acts to equalize electron and ion temperatures.

\subsubsection{Hot electron, hot ion (HEHI)}
\label{hehi section}

\begin{figure}[h!]
\centering
\includegraphics[width=1\linewidth]{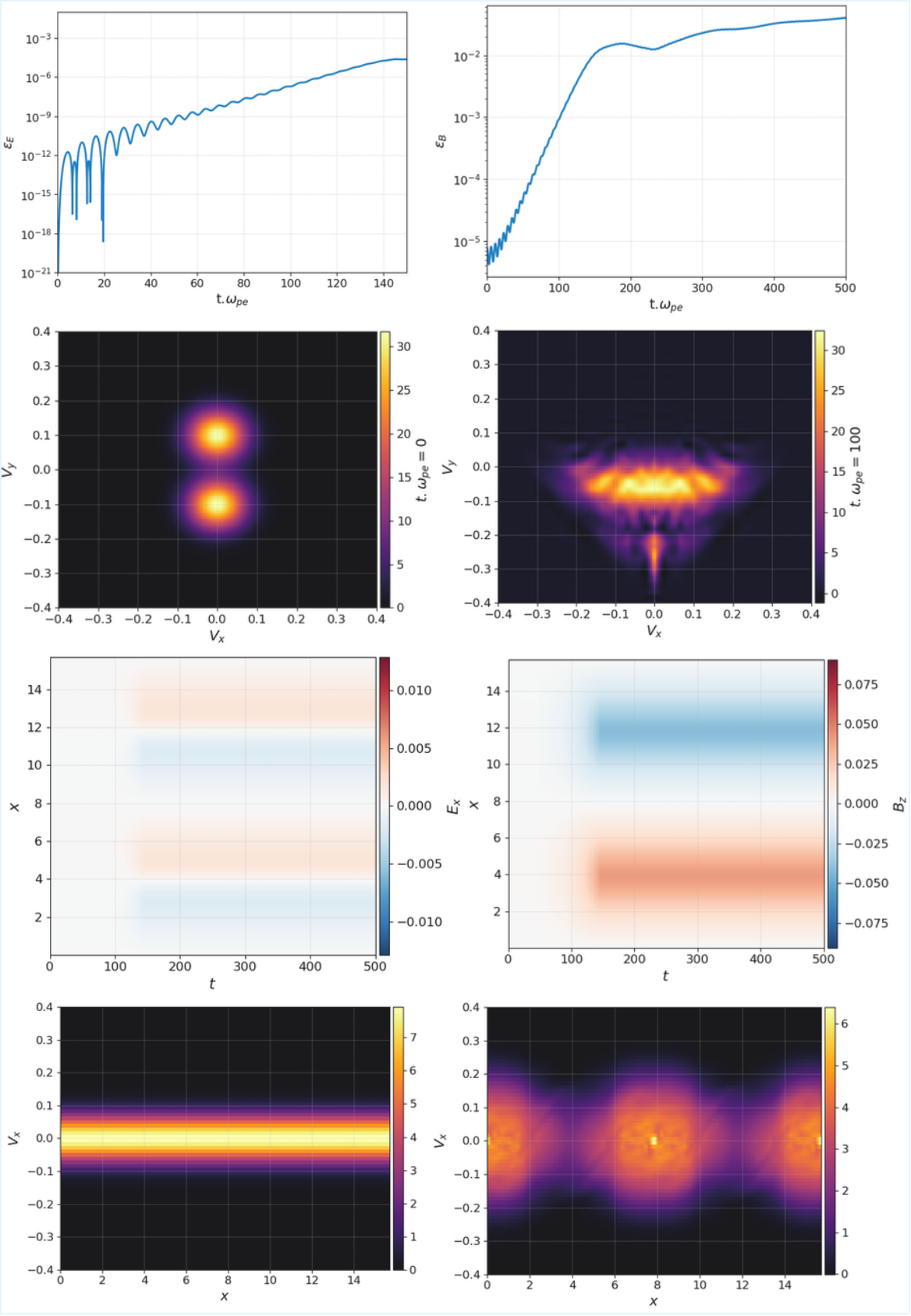}
\caption{Multi-species hot electron, hot ion (HEHI) case. \textit{Top panel}: time evolution of electric field energy (left) and magnetic field energy (right). The magnetic field energy dominates, indicating that magnetic trapping is again the dominant saturation mechanism. \textit{Second panel}: velocity distribution function in $V_x$--$V_y$ space at $t\omega_{pe}=0$ (left) and $t\omega_{pe}=100$ (right). \textit{Third panel}: spatio-temporal evolution of electric field $E_x$ (left) and magnetic field $B_z$ (right). \textit{Bottom panel}: phase-space distribution of electrons in $x$--$V_x$ at $t\omega_{pe}=0$ (left) and $t\omega_{pe}=100$ (right).}
\label{hehi}
\end{figure}

\begin{figure}[h!]
\centering
\includegraphics[width=1\linewidth]{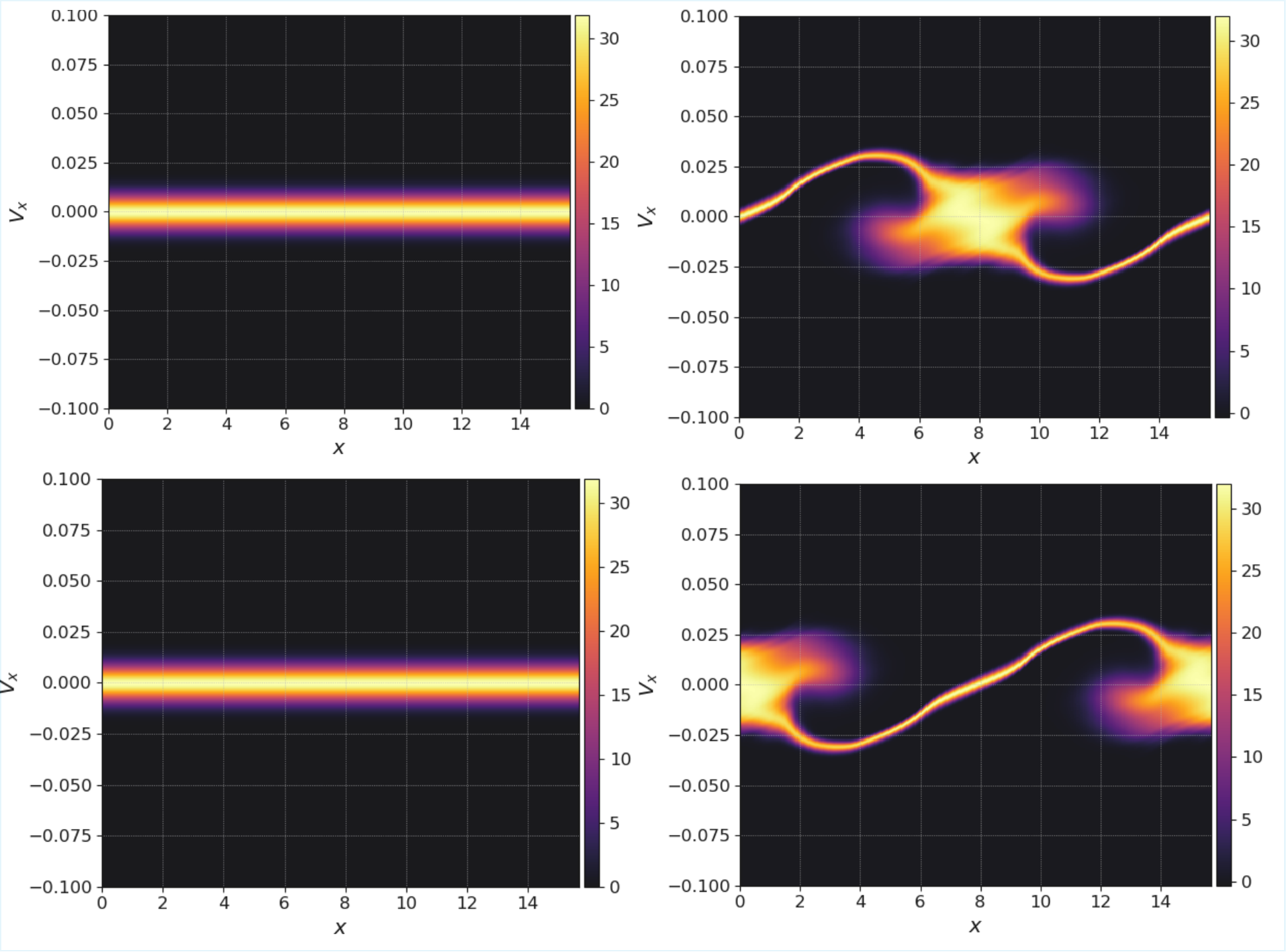}
\caption{Ion phase-space distribution for the HEHI case. \textit{Top panel}: distribution function of the first ion species (incoming from the left) in $x$--$V_x$ space at $t\omega_{pe}=0$ (left) and $t\omega_{pe}=100$ (right). \textit{Bottom panel}: same for the second ion species (incoming from the right). The ions show slow but distinct channel formation and gradual thermalization, in contrast to the rapid equilibration of the electrons.}
\label{hehi ion}
\end{figure}

Simulations for the HEHI case were performed with parameters given in Table~\ref{simulaton overview}. From the top panel of Fig.~\ref{hehi}, the electric field grows with some noise until $t\omega_{pe}=60$, then continues growing to $t\omega_{pe}=142$ before saturating. The magnetic field energy grows to $t\omega_{pe}=172$ and then saturates, slightly later than in the HECI case ($t\omega_{pe}=158$), as the additional free energy stored in the hot ions provides a secondary source of anisotropy that may sustain the instability marginally longer, though this difference is near the level of simulation noise. The growth of electric field energy is much smaller than the magnetic field energy, consistent with magnetic trapping being the dominant saturation mechanism for hot beams regardless of ion temperature. The second panel of Fig.~\ref{hehi} shows the evolution of the electron velocity distribution function in $V_x$--$V_y$ space. At $t\omega_{pe}=0$ the two beams are clearly separated; by $t\omega_{pe}=100$ they have merged into a single broadened population, consistent with rapid electron thermalization. The third panel shows the spatio-temporal evolution of $E_x$ and $B_z$, where the formation and gradual downstream spreading of filamentation structures are visible. The bottom panel shows the electron phase-space distribution in $x$--$V_x$, showing the onset of particle trapping and broadening along $V_x$. The ion phase-space evolution is shown separately in Fig.~\ref{hehi ion}. Unlike the electrons, the two ion populations remain visibly distinct even at $t\omega_{pe}=100$. Both the first species (incoming from the left, top panel) and the second species (incoming from the right, bottom panel) maintain separate, beam-like structures in $x$--$V_x$ space with only modest spatial spreading. This shows that while electrons rapidly thermalize and isotropize, ions retain their directed bulk velocities and undergo a much slower thermalization. The persistence of distinct ion velocity channels is responsible for the continued slow growth of the magnetic energy observed after the initial saturation, and for the progressive merging of ion current channels that drives the late-time dynamics described in the abstract.

\subsubsection{Cold electron, hot ion (CEHI)}
\label{cehi section}

\begin{figure}[h!]
\centering
\includegraphics[width=1\linewidth]{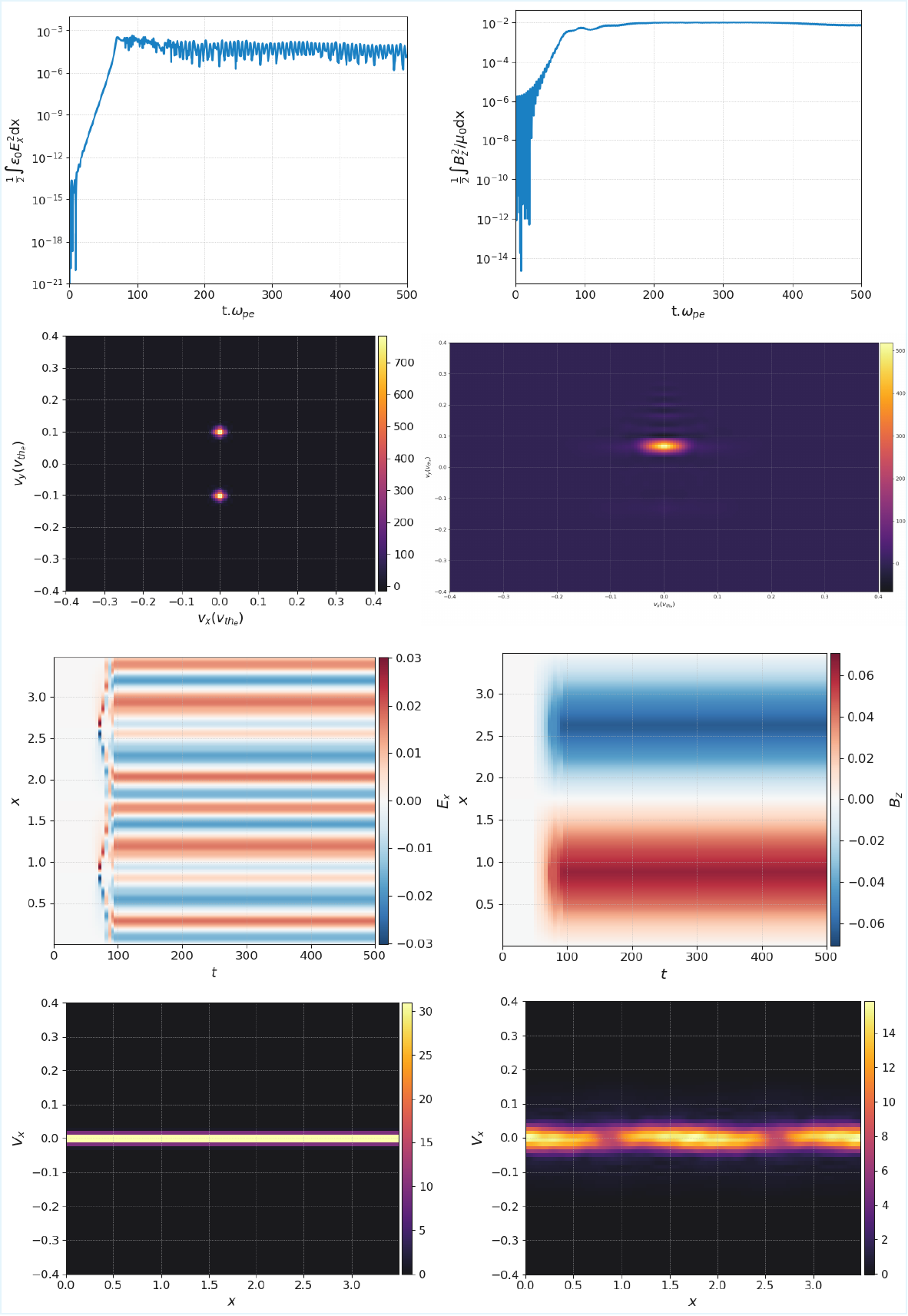}
\caption{Multi-species cold electron, hot ion (CEHI) case. \textit{Top panel}: time evolution of electric field energy (left) and magnetic field energy (right). The electric and magnetic field energies reach comparable levels at saturation, mirroring the single-species cold case and consistent with the dominant role of cold electron dynamics in the saturation behavior. \textit{Second panel}: velocity distribution function in $V_x$--$V_y$ space at $t\omega_{pe}=0$ (left) and $t\omega_{pe}=100$ (right). \textit{Third panel}: spatio-temporal evolution of electric field $E_x$ (left) and magnetic field $B_z$ (right). \textit{Bottom panel}: phase-space distribution in $x$--$V_x$ at $t\omega_{pe}=0$ (left) and $t\omega_{pe}=100$ (right).}
\label{cehi}
\end{figure}

Simulations for the CEHI case were performed with parameters given in Table~\ref{simulaton overview}. From the top panel of Fig.~\ref{cehi}, the electric field grows with some noise until $t\omega_{pe}=14$, after which the noise disappears; the field continues growing until $t\omega_{pe}\simeq70$ and then saturates. The magnetic field energy grows to $t\omega_{pe}=75$ and then saturates, notably earlier than the CECI case ($t\omega_{pe}\simeq115$), because the hot ions, having a larger thermal spread, respond more readily to the growing electromagnetic structures and contribute additional pitch-angle scattering that accelerates the approach to isotropy. The electric and magnetic field energies reach comparable saturation levels, showing the same behavior as the single-species cold case and consistent with cold electron temperature governing the saturation mechanism regardless of ion temperature. Note that for the CEHI case, Theory~I results should be interpreted with caution: the analytical approximation in Appendix~\ref{app F} requires $u_{di}/v_{thi}\ll1$, but for the CEHI parameters ($v_{thi}/u_d=0.5$, $u_d=0.1c$) one has $u_{di}/v_{thi}=2.0$, which violates this condition. Theory~II results, which solve the dispersion relation numerically without this approximation, are the appropriate reference for this case. The second panel of Fig.~\ref{cehi} shows the electron velocity distribution function at $t\omega_{pe}=0$ and $t\omega_{pe}=100$. The two initially separated electron beams merge into a single population by the time of saturation, while the hot ions, being heavier and slower to respond, retain a broader distribution due to their higher initial thermal spread. The third panel shows the spatio-temporal evolution of the electric field $E_x$ (left) and the magnetic field $B_z$ (right). The fields develop clear filamentation patterns along $x$, with the magnetic filaments saturating around $t\omega_{pe}\simeq75$, consistent with the energy evolution in the top panel. The bottom panel shows the electron phase-space distribution in $x$--$V_x$ at $t\omega_{pe}=0$ and $t\omega_{pe}=100$, illustrating particle trapping and the broadening of the distribution function due to phase mixing. The overall saturation behavior in this case is dominated by the cold electrons: the electrostatic potential wells and magnetic trapping contribute in comparable measure, just as in the single-species cold case.

\section{Observational Context}
\label{observations}

The counter-streaming beam configurations studied in the preceding sections arise naturally in collisionless astrophysical plasmas, and most directly at the foot of quasi-perpendicular shocks where reflected ions form a beam propagating against the incident solar wind. To place the simulation parameter space within directly measurable plasma environments, we present two observational figures. The first compares the simulation parameters to the bulk solar wind distribution in the temperature anisotropy plane. The second shows a magnetospheric bow shock crossing that exhibits differential electron-ion thermalization qualitatively consistent with our simulation results.

\subsection{Temperature anisotropy in the solar wind (Wind/SWE)}
\label{obs wind}

\begin{figure}[h!]
\centering
\includegraphics[width=1\linewidth]{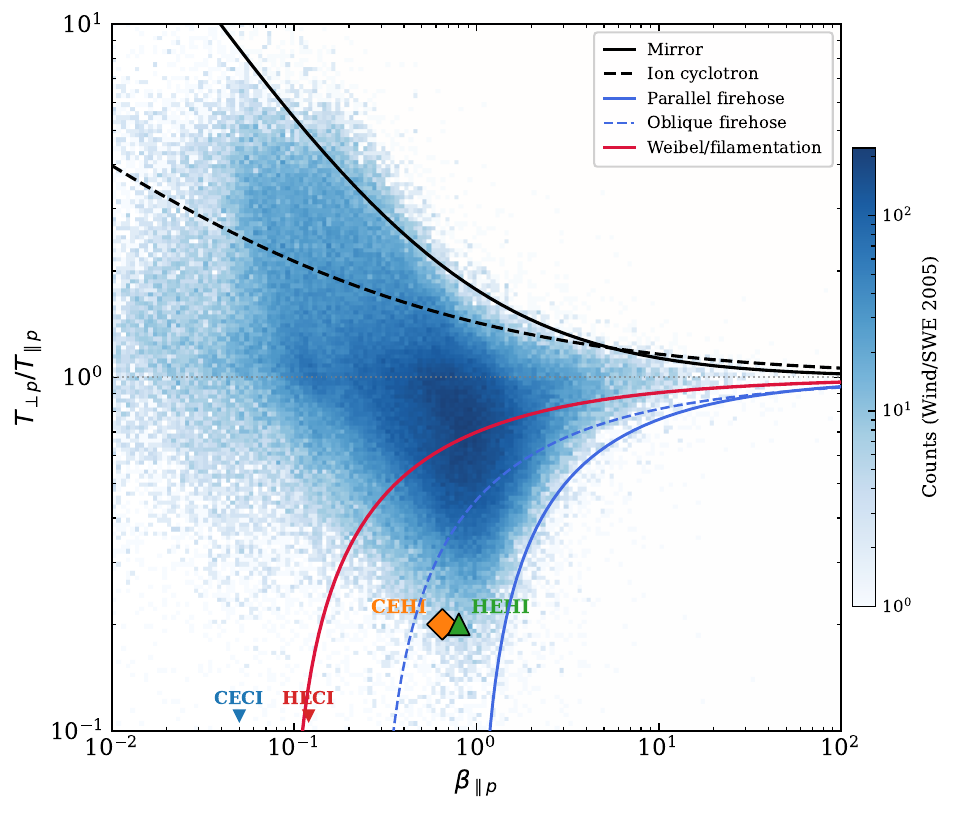}
\caption{Proton temperature anisotropy from Wind/SWE observations during 2005 ($\approx3.4\times10^5$ measurements) in the $\beta_{\parallel p}$--$T_{\perp p}/T_{\parallel p}$ plane. Instability threshold curves follow Hellinger \textit{et al.}~\cite{hellinger2006solar}; the Weibel/filamentation onset is in red. Markers show $A_{\mathrm{eff}}^{(i)}$ (equation~\ref{A eff}) for each simulation case; downward carets (CECI, HECI) indicate $A_{\mathrm{eff}} \approx 0.01$, below the plotted range. All four cases lie in the firehose/Weibel-unstable region ($T_\perp < T_\parallel$).}
\label{obs brazil}
\end{figure}

Figure~\ref{obs brazil} shows the proton temperature anisotropy distribution from approximately $3.4\times10^5$ Wind spacecraft Solar Wind Experiment (SWE) measurements recorded during the calendar year 2005~\cite{ogilvie1995swe}, displayed in the $\beta_{\parallel p}$--$T_{\perp p}/T_{\parallel p}$ plane, often called the Brazil plot~\cite{bale2009magnetic}. The year 2005 falls within solar cycle 23 near solar minimum, when the solar wind is relatively steady and the proton distribution is well-sampled; results are qualitatively unchanged for other years in the Wind/SWE archive. Parallel and perpendicular proton temperatures are derived from the nonlinear bi-Maxwellian fit to the ion distribution function, and $\beta_{\parallel p} = 2\mu_0 n k_B T_{\parallel p}/B^2$ is computed using the magnetic field magnitude from the MFI fluxgate magnetometer~\cite{lepping1995wind}, interpolated onto the SWE timestamps. The Wind data populate the central region of the diagram, bounded above by the mirror and ion cyclotron (EMIC) thresholds and below by the parallel and oblique firehose thresholds; the curves plotted follow the parametric fits of Hellinger \textit{et al.}~\cite{hellinger2006solar}. The clustering of measurements near these threshold curves is broadly consistent with regulation of temperature anisotropy by kinetic instabilities. The Weibel/filamentation threshold follows the approximate form $T_\perp/T_\parallel \approx 1 - 0.30\,\beta^{-0.5}$~\cite{gary1996whistler, skoutnev2019temperature} and marks the onset of the filamentation instability below the firehose threshold. The four simulation cases are placed at their effective ion anisotropy in the beam frame,

\begin{equation}
A_{\mathrm{eff}}^{(i)} = \frac{v_{thi}^2}{v_{thi}^2 + u_d^2},
\label{A eff}
\end{equation}

where the denominator accounts for both thermal and directed kinetic energy along the beam direction. Because $u_d/c = 0.1$ in all cases and the thermal-to-drift ratio $v_{thi}/u_d$ is $0.1$ for the cold-ion cases (CECI, HECI) and $0.5$ for the hot-ion cases (HEHI, CEHI), equation~(\ref{A eff}) gives $A_{\mathrm{eff}}^{(i)} \approx 0.010$ and $0.20$ respectively. All four simulation initial conditions therefore lie in the firehose/Weibel-unstable region ($T_\perp < T_\parallel$), well below the isotropic line, indicating that they represent strongly anisotropic beam configurations. The schematic $\beta$ values are chosen to represent the post-shock magnetosheath environment ($0.05 \lesssim \beta_{\parallel p} \lesssim 1$). The simulations start unmagnetized ($B_0\to0$), so the initial $\beta$ is formally infinite; the marker positions therefore indicate the relevant qualitative regime rather than an exact initial coordinate.

\subsection{MMS bow shock crossing (differential thermalization)}
\label{obs mms}

\begin{figure}[h!]
\centering
\includegraphics[width=1\linewidth]{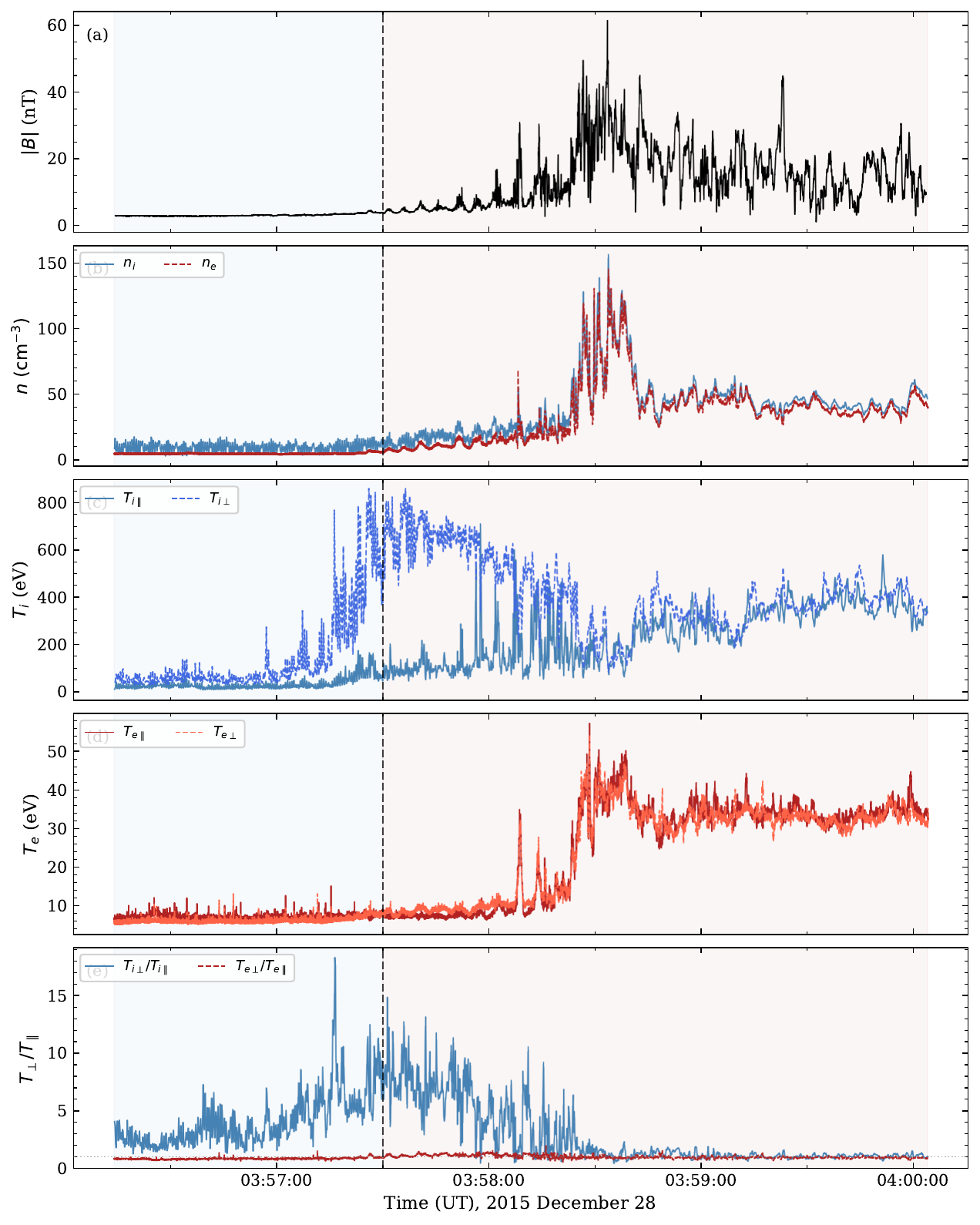}
\caption{MMS1 quasi-perpendicular bow shock crossing on 2015 December~28 ($\theta_{Bn}\approx83^\circ$, $M_A\approx27$; Madanian \textit{et al.}~\cite{madanian2021dynamics}). Dashed line: shock ramp; blue (red) shading: upstream (downstream). (a)~$|B|$ (FGM). (b)~Ion and electron densities (FPI). (c)~Ion $T_\parallel$, $T_\perp$ (FPI DIS). (d)~Electron $T_\parallel$, $T_\perp$ (FPI DES). (e)~$T_\perp/T_\parallel$ for ions (blue) and electrons (red); dotted line marks isotropy. Ions reach $T_{i\perp}/T_{i\parallel} \sim5$--$15$ across the ramp and relax slowly downstream, while electrons remain near isotropy throughout.}
\label{fig:mms}
\end{figure}

Figure~\ref{fig:mms} shows MMS1 observations of a quasi-perpendicular Earth bow shock crossing on 2015 December~28 near 03:57~UT, studied in detail by Madanian \textit{et al.}~\cite{madanian2021dynamics}. The shock geometry is $\theta_{Bn}\approx83^\circ$ with an Alfv\'en Mach number $M_A\approx27$. This places the event in a regime where kinetic instabilities may contribute to the shock-foot structure~\cite{juno:2021}, although the present data do not by themselves uniquely identify the dominant microphysical mechanism. The magnetic field magnitude is from the Flux Gate Magnetometer (FGM; burst mode, 128~vectors\,s$^{-1}$)~\cite{russell2016magnetospheric}, and the particle moments are from the Fast Plasma Investigation (FPI; burst mode, 150~ms ions and 30~ms electrons)~\cite{pollock2016fast}. Panel~(a) shows the magnetic field compression across the shock ramp, with $|B|$ rising from about 5~nT upstream to 20--60~nT downstream. Panel~(b) shows the corresponding increase in ion and electron densities, confirming the shock transition. Panels~(c) and (d) give the parallel and perpendicular temperature components for ions and electrons; ions are heated from about 50~eV to 400~eV while electrons are heated from about 7~eV to 35~eV.

The key observational feature is shown in panel~(e): the electron anisotropy $T_{e\perp}/T_{e\parallel}$ remains close to unity throughout the transition, while the ion anisotropy $T_{i\perp}/T_{i\parallel}$ reaches values of $5$ to $15$ across the ramp and relaxes toward unity only gradually downstream. This indicates that electrons approach isotropy more rapidly than ions across the shock transition. Such an ordering is qualitatively consistent with the behavior observed in the simulations, in which electron thermalization occurs on shorter timescales than ion thermalization across a range of temperature conditions. However, the MMS observations alone do not uniquely establish the underlying microphysical mechanism, which may involve a combination of wave-particle interactions and filamentation-related processes.

The upstream counter-streaming geometry relevant to the simulations is established naturally at the shock foot. A fraction of the incoming solar wind protons are reflected and propagate upstream, forming a counter-streaming ion population with $n_b/n_0\sim 0.1$--$0.3$~\cite{sckopke1983evolution}. This configuration is analogous to the initial conditions used in the simulations. However, significant caution is warranted when comparing the simulation results directly to this particular event. In high-Mach-number quasi-perpendicular shocks, the shock reformation cycle operates on a typical timescale of $\sim 1.5\,\Omega_i^{-1}$, during which the Weibel instability appears and disappears periodically~\cite{bohdan2020kinetic, bohdan2021magnetic}. Within each reformation cycle, the effective window available for Weibel growth is approximately $\sim 1\,\Omega_i^{-1} \approx 82\,\omega_{pe}^{-1}$ at the parameters of this event. Whether the instability reaches its nonlinear stage within this interval depends on the shock foot being sufficiently extended relative to the linear growth length~\cite{bohdan2020kinetic}; for the $M_A \approx 27$ conditions here, this condition may not be comfortably satisfied. The present shock crossing spans approximately $20\,\Omega_i^{-1}$ in duration, which encompasses roughly 13 such reformation cycles. Accordingly, this event should be viewed as a time-integrated observation spanning many reformation phases rather than as a direct window onto a single nonlinear Weibel growth stage, and the comparison with the simulation results should be treated as qualitative and order-of-magnitude at best.

The simulation drift velocity $u_d/c = 0.1$ corresponds to high-energy astrophysical environments such as laser-plasma systems and young supernova remnants, whereas Earth's bow shock has $u_d/c\sim10^{-3}$. Consequently, the instability growth timescales measured in $\omega_{pe}^{-1}$ differ substantially between the two systems because $\omega_{pe}^{-1}$ itself scales with the plasma density and the shock speed. However, as noted in Bohdan \textit{et al.}~\cite{bohdan2020kinetic, bohdan2021magnetic}, when growth timescales are expressed in units of $\Omega_i^{-1}$ rather than $\omega_{pe}^{-1}$, the normalization is the same across simulations and physical bow shocks, and direct comparison is physically meaningful. The qualitative ordering of electron and ion thermalization observed here is consistent with the trends obtained in the simulations when viewed in this common dimensionless frame. Nevertheless, as discussed above, the limited Weibel growth window within each reformation cycle introduces genuine uncertainty as to whether the nonlinear Weibel stage is reached in the present event; the observational data therefore provide qualitative support for the physical picture explored in this study rather than a direct quantitative validation.

\section{Summary \& Conclusions}
\label{Summary}

This study investigated the Weibel instability in the presence of non-stationary ions. We used 1X2V continuum Vlasov-Maxwell simulations of interpenetrating plasma flows, incorporating a non-stationary ion background. We used three approaches to calculate growth rates: analytical, numerical, and direct simulation. For the analytical part, the plasma dispersion function was expanded using asymptotic and power series expansions. For the numerical solution, we employed an optimization method using SciPy's \texttt{fsolve} to locate the roots of the dispersion function corresponding to each wavenumber, starting from an initial estimate and iteratively refining until convergence. For the simulation, the full Vlasov-Maxwell equations were solved using the \texttt{Gkeyll} framework. Using these three approaches, we discussed the ion-Weibel instability alongside the scenario involving only electrons. Two single-species cases were examined: hot and cold electron beams. In the hot electron case, kinetic energy was converted significantly into magnetic field energy, forming magnetic filaments that coalesced and trapped particles. The electric field energy remained negligibly small compared to the magnetic field energy, so magnetic trapping dominated the Weibel saturation. In the cold electron case, the electric field energy grew to levels comparable to the magnetic field energy. The resulting electric potential wells played an equal role in saturating the instability. For the multi-species cases, we considered ions as a non-stationary background and examined four temperature combinations. The electron species reacted rapidly to the fields while ions responded slowly due to their inertia. In hot-electron cases (HECI and HEHI), magnetic trapping remained the dominant saturation mechanism regardless of ion temperature. In cold-electron cases (CECI and CEHI), the electric and magnetic fields reached comparable energies, mirroring the single-species cold behavior and consistent with the electron temperature governing the saturation mechanism. The instability saturated in all cases as the free energy stored in the beam anisotropy was converted into electric and magnetic field energies. Table~\ref{tab:summary} provides a concise comparison of the saturation timescales and dominant mechanisms across all six cases.

\begin{table}[h!]
\caption{Summary of saturation timescales and dominant mechanisms for all cases studied. The saturation time $t_\mathrm{sat}$ is given in units of $\omega_{pe}^{-1}$. The saturation mechanism is classified by comparing the peak electric and magnetic field energies at saturation: cases where the peak magnetic field energy exceeds the peak electric field energy by more than one order of magnitude are classified as magnetically dominated (MT); cases where the two peak energies are within one order of magnitude involve both electrostatic and magnetic trapping (ES+MT).}
\begin{ruledtabular}
\begin{tabular}{cccc}
Case & $t_\mathrm{sat}\,\omega_{pe}$ & Field energies & Mechanism \\
\hline
HE   & $\sim100$ & $B\gg E$ & MT \\
CE   & $\sim60$  & $B\approx E$ & ES+MT \\
CECI & $\sim115$ & $B\approx E$ & ES+MT \\
HECI & $\sim158$ & $B\gg E$ & MT \\
HEHI & $\sim172$ & $B\gg E$ & MT \\
CEHI & $\sim75$  & $B\approx E$ & ES+MT \\
\end{tabular}
\end{ruledtabular}
\label{tab:summary}
\end{table}

The table reveals a clear organizing principle: the saturation mechanism, whether dominated by magnetic trapping or by comparable electric and magnetic fields, is determined entirely by the electron temperature and is insensitive to the ion temperature. Cold-electron cases all exhibit comparable electric and magnetic energies at saturation, while hot-electron cases are universally dominated by magnetic trapping. Furthermore, in the hot-electron multi-species cases (HECI and HEHI), the magnetic energy continues to grow slowly after initial saturation due to the persistence of distinct ion velocity channels, as seen in the ion phase-space analysis (Fig.~\ref{hehi ion}). These ion channels merge at late times and drive a secondary, slower phase of field amplification. These results are relevant to collisionless shock formation in astrophysical compact objects and laboratory laser-plasma experiments, where interpenetrating plasma streams of varying temperature are ubiquitous. The finding that ion dynamics, while slow, ultimately alter the late-time saturation behavior through the formation and merging of shielded ion current channels supports the importance of treating ions as a fully kinetic, evolving species even when electron-driven growth dominates the linear phase. Section~\ref{observations} places these findings in an observational context. Using approximately $3.4\times10^5$ Wind/SWE proton measurements recorded during 2005, we showed that all four simulation initial conditions sit in the firehose/Weibel-unstable region of the $\beta_{\parallel p}$--$T_{\perp p}/T_{\parallel p}$ diagram (Fig.~\ref{obs brazil}), below the Weibel/filamentation threshold and well below the isotropic line. The concentration of solar wind measurements near the instability threshold curves shows that the anisotropy regime explored in the simulations is represented in ambient solar wind measurements. The MMS1 quasi-perpendicular bow shock crossing of 2015 December~28 (Fig.~\ref{fig:mms}) provides an observational comparison for the central simulation result. Panel~(e) of that figure shows that $T_{i\perp}/T_{i\parallel}$ reaches values of $5$ to $15$ across the shock ramp and relaxes toward unity only gradually in the downstream magnetosheath, while $T_{e\perp}/T_{e\parallel}$ remains close to unity throughout the entire transition. This is qualitatively consistent with the simulation result that electrons reach thermal equilibrium rapidly through magnetic trapping, while ions retain distinct bulk velocities within shielded current channels and thermalize on a much longer timescale. Taken together, both observational comparisons are consistent with the ordering of saturation and thermalization timescales established in the simulations, and provide qualitative support for the view that this ordering is physically realized in natural collisionless shock environments. A direct quantitative comparison, however, is subject to the caveats discussed in Section~\ref{obs mms}, particularly the limited Weibel growth window per reformation cycle at Earth's bow shock. The present simulations are carried out in one spatial dimension (1X2V). This geometry reproduces the linear growth and early nonlinear saturation of the Weibel instability, and the ordering of electron and ion thermalization timescales is not expected to change with dimensionality. However, the 1D geometry precludes oblique filamentation modes and the full two-dimensional coalescence dynamics of current filaments, which can modify the late-time saturation amplitude and the rate of ion channel merging. Extending the present study to a 2X2V geometry is a natural next step and would allow direct comparison with two-dimensional PIC results in the literature.

\begin{acknowledgments}
VS and RM sincerely thank James Juno and Ammar Hakim at Princeton Plasma Physics Laboratory, and Petr Cagas at Virginia Tech for their valuable suggestions and insightful discussions. Simulations were performed at the High-Performance Computing Centre \textit{Brahmagupta} at Sikkim University and on the \textit{Stellar} cluster at Princeton University. We gratefully acknowledge the entire \texttt{Gkeyll} team for their continuous support and development of the code. The \texttt{Gkeyll} framework is publicly available and can be downloaded from \href{https://github.com/ammarhakim/gkyl}{GitHub}. The code and input files used in this work are accessible through the \texttt{Gkeyll} repository, and the results are reproducible using \texttt{Gkeyll}~2.0. Detailed installation and usage instructions can be found at \url{https://gkeyll.readthedocs.io/en/latest/}.
\end{acknowledgments}

\section*{Author Contributions}
V.S. developed the simulation framework, performed the numerical simulations, derived the analytical growth rates, and led the writing of the manuscript. M.K.C. carried out the full observational data analysis, contributed to the formal analysis, and wrote the corresponding sections. H.D.S. contributed to the formal analysis and reviewed the manuscript. B.S. provided supporting contributions and reviewed the manuscript. R.M. supervised the study throughout and contributed to the formal analysis. All authors reviewed and approved the final manuscript.

\section*{Data Availability}
Wind/SWE and MMS data are publicly available from the NASA CDAWeb archive (\url{https://cdaweb.gsfc.nasa.gov}) and the MMS Science Data
Center (\url{https://lasp.colorado.edu/mms/sdc/}), respectively. Simulation input files are available through the \texttt{Gkeyll} repository at \url{https://github.com/ammarhakim/gkyl}.

\bibliography{References}

\newpage
\appendix

\begin{widetext}

\section{Discontinuous Galerkin scheme and energy conservation}
\label{dg scheme appendix}

The discontinuous Galerkin (DG) method combines the accuracy of the finite element method with the stability properties of the finite volume method. It uses high-order polynomials for accuracy and localizes data for efficient parallelization. Here we discuss the DG discretization and the conservation of total energy.

\begin{equation}
\frac{\partial f_s}{\partial t}+\Vec{v}\cdot\Vec{\nabla}_x f_s
+ \frac{q_s}{m_s}(\Vec{E}+\Vec{v}\times \Vec{B})\cdot \Vec{\nabla}_v f_s=0
\label{dg eq}
\end{equation}

Integrating equation~(\ref{dg eq}) over all of phase space and summing over all species,

\begin{equation}
\frac{d}{dt}\int_{\Omega}\sum_s \left\langle \frac{\partial f_s}{\partial t}
+\Vec{v}\cdot\Vec{\nabla}_x f_s
+ \frac{q_s}{m_s}(\Vec{E}+\Vec{v}\times \Vec{B})\cdot \Vec{\nabla}_v f_s\right\rangle dx=0.
\end{equation}

This shows that the Vlasov-Maxwell system is a conservation law in phase space. Multiplying equation~(\ref{dg eq}) by $m_s|\Vec{v}|^2/2$, integrating over all phase space, and summing over species,

\begin{equation}
\frac{\partial}{\partial t}\int_\Omega \sum_s\left\langle\frac{1}{2}m_s|\Vec{v}|^2\right\rangle_s dx
+\int_\Omega\Vec{\nabla}\cdot\sum_s \left\langle\frac{1}{2}m_s|\Vec{v}|^2\Vec{v}\right\rangle_s dx
-\int_\Omega \Vec{E}\cdot\Vec{J}\,dx=0,
\end{equation}

Using Poynting's theorem, $\int_\Omega \Vec{E}\cdot\Vec{J}\,dx = -\frac{d}{dt}\int_\Omega\frac{\epsilon_0 E^2 + B^2/\mu_0}{2}\,dx$, which gives the total energy conservation law,

\begin{equation}
\frac{d}{dt}\left[\int_\Omega\sum_s\frac{1}{2}m_s\langle v^2\rangle_s\,dx
+ \int_\Omega\frac{\epsilon_0 E^2}{2}\,dx + \int_\Omega\frac{B^2}{2\mu_0}\,dx\right] = 0.
\label{energy conservation appendix}
\end{equation}

\section{Linearization of Vlasov equation}
\label{app A}

We linearize the Vlasov equation~(\ref{VMeq}) by writing $f_s = f_{s,0} + f_{s,1}$, $\Vec{E} = \Vec{E}_1$, and $\Vec{B} = \Vec{B}_1$, where subscripts $0$ and $1$ denote the equilibrium and perturbation quantities respectively. Assuming perturbations of the form $e^{i(kx - \omega t)}$, the linearized Vlasov equation gives equation~(\ref{linearized_max}) as shown in Section~\ref{linear theory}.

\section{Derivation of the dispersion relation}
\label{app B}

Starting from the linearized distribution function~(\ref{linearized_max}) and substituting into the linearized Amp\`{e}re's law~(\ref{linearized_amperes_law}), we integrate over velocity space using the equilibrium distribution~(\ref{maxwellian}). After carrying out the velocity integrals and using the definition of the plasma dispersion function~(\ref{plasma dispersion function}), we arrive at the dispersion relation~(\ref{disp}). The key intermediate step involves evaluating integrals of the form

\begin{equation}
\int_{-\infty}^{\infty} \frac{v_y^2}{(\omega - v_x k_x)} \frac{\partial f_{s,0}}{\partial v_y}\,dv_y,
\end{equation}

which, after integration by parts and use of the plasma dispersion function, yields the terms proportional to $\zeta_s Z(\zeta_s)$ appearing in equation~(\ref{disp}). The full algebra is standard and follows the treatment in Appendix~B of Skoutnev \textit{et al.}~\cite{skoutnev2019temperature}. The velocity integrations yield an intermediate result of the form

\begin{equation}
1 + \frac{\omega_{pe}^2}{k_x^2 c^2}\left[\frac{u_{de}^2}{v_{the}^2}
+ \zeta_{pe}Z(\zeta_{pe})\left(1+\frac{u_{de}^2}{v_{the}^2}\right)\right]
+\frac{\omega_{pi}^2}{c^2 k_x^2} \left[-Z'(\zeta_{pi})\frac{u_d^2}{v^2_{thi}}
+2\zeta_{pi} Z(\zeta_{pi})\right]
+\frac{\omega^2}{c^2 k_x^2} = 0.
\label{finalterm}
\end{equation}

To convert equation~(\ref{finalterm}) to the form in equation~(\ref{disp}), we apply the identity $Z'(\zeta) = -2[1+\zeta Z(\zeta)]$, which gives $-Z'(\zeta_{pi}) = 2[1+\zeta_{pi}Z(\zeta_{pi})]$. Substituting into the ion term:

\begin{align}
-Z'(\zeta_{pi})\frac{u_d^2}{v^2_{thi}}+2\zeta_{pi} Z(\zeta_{pi})
&= 2[1+\zeta_{pi}Z(\zeta_{pi})]\frac{u_{di}^2}{v^2_{thi}}+2\zeta_{pi} Z(\zeta_{pi}) \nonumber\\
&= 2\frac{u_{di}^2}{v^2_{thi}} + 2\zeta_{pi}Z(\zeta_{pi})\left(1+\frac{u_{di}^2}{v^2_{thi}}\right).
\end{align}

Inserting this result into equation~(\ref{finalterm}) and changing the overall sign (since the equation equals zero) reproduces equation~(\ref{disp}) exactly.

\section{Growth rate for cold electrons and cold ions (CECI)}
\label{app C}

For cold electrons, $v_{the}$ is small and $\zeta_e$ is large. Using the asymptotic expansion $Z(\zeta_e)\sim -\frac{1}{\zeta_e}-\frac{1}{2\zeta^3_e}$, and similarly for cold ions $Z(\zeta_i)\sim -\frac{1}{\zeta_i}-\frac{1}{2\zeta^3_i}$, substituting into equation~(\ref{disp}) and simplifying gives

\begin{equation}
\omega^4-\omega^2(k^2c^2 + \omega^2_{pe}+\omega^2_{pi})
- \omega^2_{pe}u^2_{de}k^2-\omega^2_{pi}u^2_{di}k^2
-\omega^2_{pe}k^2v^2_{the}-\omega^2_{pi}k^2v^2_{thi} = 0.
\label{ceci dis}
\end{equation}

This has the form $\omega^4 - A\omega^2 - B = 0$ with $A,B>0$. The unstable root satisfies $\omega^2 = (A - \sqrt{A^2+4B})/2 < 0$, so $\omega = i\gamma$ with $\gamma^2 = (\sqrt{A^2+4B}-A)/2$. Solving for the growth rate:

\begin{equation}
\gamma=\sqrt{\frac{1}{2}\!\left[\sqrt{(k^2c^2+\omega_{pe}^2+\omega_{pi}^2)^2
+4(\omega_{pe}^2u_{de}^2k^2+\omega_{pi}^2u_{di}^2k^2
+\omega^2_{pe}k^2v^2_{the}+\omega^2_{pi}k^2v^2_{thi})}
-(k^2c^2+\omega_{pe}^2+\omega_{pi}^2)\right]}.
\label{cecianalyappendix}
\end{equation}

\section{Growth rate for hot electrons and cold ions (HECI)}
\label{app D}

For hot electrons, $\zeta_e$ is small, so $Z(\zeta_e)\sim -2\zeta_e + \frac{4\zeta^3_e}{3}$. For cold ions, $\zeta_i$ is large, so $Z(\zeta_i)\sim -\frac{1}{\zeta_i}-\frac{1}{2\zeta^3_i}$. Substituting into equation~(\ref{disp}) and simplifying:

\begin{equation}
\omega^4\!\left(1-\frac{\omega^2_{pe}}{v_{the}^2k^2}
-\frac{\omega^2_{pe}u_{de}^2}{v_{the}^4k^2}\right)
-\omega^2\!\left(k^2c^2-\frac{\omega^2_{pe}u_{de}^2}{v_{the}^2}+\omega^2_{pi}\right)
-\left(\omega^2_{pi}v^2_{thi}k^2+\omega^2_{pi}u^2_{di}k^2\right)=0.
\end{equation}

The growth rate is

\begin{equation}
\gamma= \sqrt{\frac{A_\mathrm{H} - \sqrt{A_\mathrm{H}^2
+4\,C_\mathrm{H}\,D_\mathrm{H}}}{2\,C_\mathrm{H}}},
\label{heci dispersion relation app}
\end{equation}

where

\begin{align}
A_\mathrm{H} &= k^2c^2-\frac{\omega^2_{pe}u^2_{de}}{v_{the}^2}+\omega_{pi}^2, \nonumber\\
C_\mathrm{H} &= 1-\frac{\omega^2_{pe}}{v_{the}^2k^2}
-\frac{\omega^2_{pe}u_{de}^2}{v_{the}^4k^2}, \nonumber\\
D_\mathrm{H} &= \omega^2_{pi}v^2_{thi}k^2+\omega^2_{pi}u^2_{di}k^2.
\nonumber
\end{align}

The radicand in equation~(\ref{heci dispersion relation app}) is positive in the unstable regime. At the relevant wavenumbers for hot electrons, $\omega_{pe}^2(1+u_{de}^2/v_{the}^2)/(v_{the}^2k^2) \gg 1$, so $C_\mathrm{H}<0$ and $D_\mathrm{H}>0$, giving $4C_\mathrm{H}D_\mathrm{H}<0$ and hence $A_\mathrm{H}^2+4C_\mathrm{H}D_\mathrm{H}<A_\mathrm{H}^2$. The instability condition $A_\mathrm{H}<0$ (i.e.\ $\omega_{pe}^2u_{de}^2/v_{the}^2>k^2c^2+\omega_{pi}^2$) then ensures that $A_\mathrm{H}-\sqrt{A_\mathrm{H}^2+4C_\mathrm{H}D_\mathrm{H}}<0$, so the numerator and $C_\mathrm{H}$ carry the same sign and the overall radicand is positive.

\section{Growth rate for hot electrons and hot ions (HEHI)}
\label{app E}

For hot electrons and hot ions, both $\zeta_e$ and $\zeta_i$ are small. Using the power series expansion for both species: $Z(\zeta_e)\sim -2\zeta_e + \frac{4\zeta^3_e}{3}$ and $Z(\zeta_i)\sim -2\zeta_i + \frac{4\zeta^3_i}{3}$. Substituting into equation~(\ref{disp}) and simplifying:

\begin{equation}
\begin{split}
& \omega^4\!\left(\frac{\omega_{pe}^2}{3v^4_{the}k^4}
+\frac{\omega_{pe}^2 u_{de}^2}{3v^6_{the}k^4}
+\frac{\omega_{pi}^2}{3v^4_{thi}k^4}
+\frac{\omega_{pi}^2 u_{di}^2}{3v^6_{thi}k^4}\right)\\
&-\omega^2\!\left(-1+\frac{\omega_{pe}^2}{v^2_{the}k^2}
+\frac{\omega_{pe}^2u_{de}^2}{v^4_{the}k^2}
+\frac{\omega_{pi}^2}{v^2_{thi}k^2}
+\frac{\omega_{pi}^2u_{di}^2}{v^4_{thi}k^2}\right)\\
&+\left(-k^2c^2+\frac{\omega_{pe}^2u_{de}^2}{v^2_{the}}
+\frac{\omega_{pi}^2u_{di}^2}{v^2_{thi}}\right)=0.
\end{split}
\label{hehi disp rel appendix}
\end{equation}

The growth rate is

\begin{equation*}
\resizebox{1.0\linewidth}{!}{$
\gamma = \sqrt{
\frac{
\left(-1+\frac{\omega_{pe}^2}{v^2_{the}k^2}+\frac{\omega_{pe}^2u_{de}^2}{v^4_{the}k^2}
+\frac{\omega_{pi}^2}{v^2_{thi}k^2}+\frac{\omega_{pi}^2u_{di}^2}{v^4_{thi}k^2}\right)
-\sqrt{
\left(-1+\frac{\omega_{pe}^2}{v^2_{the}k^2}+\frac{\omega_{pe}^2u_{de}^2}{v^4_{the}k^2}
+\frac{\omega_{pi}^2}{v^2_{thi}k^2}+\frac{\omega_{pi}^2u_{di}^2}{v^4_{thi}k^2}\right)^2
- 4\left(\frac{\omega_{pe}^2}{3v^4_{the}k^4}+\frac{\omega_{pe}^2u_{de}^2}{3v^6_{the}k^4}
+\frac{\omega_{pi}^2}{3v^4_{thi}k^4}+\frac{\omega_{pi}^2u_{di}^2}{3v^6_{thi}k^4}\right)
\left(-k^2c^2+\frac{\omega_{pe}^2u_{de}^2}{v^2_{the}}
+\frac{\omega_{pi}^2u_{di}^2}{v^2_{thi}}\right)}
}{
2\left(\frac{\omega_{pe}^2}{3v^4_{the}k^4}+\frac{\omega_{pe}^2u_{de}^2}{3v^6_{the}k^4}
+\frac{\omega_{pi}^2}{3v^4_{thi}k^4}+\frac{\omega_{pi}^2u_{di}^2}{3v^6_{thi}k^4}\right)
}
}$}
\end{equation*}

\section{Growth rate for cold electrons and hot ions (CEHI)}
\label{app F}

For hot ions, $\zeta_i$ is small, so $Z(\zeta_i)\sim -2\zeta_i + \frac{4\zeta^3_i}{3}$. For cold electrons, $\zeta_e$ is large, so $Z(\zeta_e)\sim -\frac{1}{\zeta_e}-\frac{1}{2\zeta^3_e}$. Substituting into equation~(\ref{disp}), multiplying by $\omega^4$, and simplifying yields a sixth-order equation in $\omega$:

\begin{equation}
\omega^6\!\left(\frac{\omega^2_{pi}}{3v_{thi}^4k^2}+\frac{\omega^2_{pi}u_{di}^2}{3v_{thi}^6k^4}\right)
+\omega^4\!\left(1-\frac{\omega^2_{pi}}{v_{thi}^2k^2}-\frac{\omega^2_{pi}u_{di}^2}{v_{thi}^4k^2}\right)
-\omega^2\!\left(k^2c^2-\frac{\omega^2_{pi}u_{di}^2}{v_{thi}^2}+\omega^2_{pe}\right)
-\left(\omega^2_{pe}v^2_{the}k^2+\omega^2_{pe}u^2_{de}k^2\right)=0.
\label{cehi sixth order}
\end{equation}

\textit{Caveat:} The following reduction from sixth to second order in $\omega^2$ assumes $u_{di}/v_{thi}\ll1$, which makes the $\omega^6$ coefficient negligibly small. However, for the CEHI simulation parameters ($v_{thi}/u_d=0.5$, $u_d=0.1c$), one has $u_{di}/v_{thi}=2.0$, which \emph{violates} this condition. The growth rate expression derived below (Theory~I for CEHI) should therefore be treated as an approximation of limited accuracy; Theory~II, which numerically solves the full dispersion relation~(\ref{disp}) without this approximation, provides the reliable quantitative reference for this case. In the regime where $v_{thi}$ is large and $u_{di}/v_{thi}\ll1$ (not satisfied for the CEHI parameters, as noted above), the $\omega^6$ coefficient is negligible compared to the $\omega^4$ coefficient, so equation~(\ref{cehi sixth order}) reduces to a quadratic in $\omega^2$, yielding the approximate growth rate:

\begin{equation}
\gamma= \sqrt{\frac{A_\mathrm{C} - \sqrt{A_\mathrm{C}^2
+4\,C_\mathrm{C}\,D_\mathrm{C}}}{2\,C_\mathrm{C}}},
\end{equation}

where

\begin{align}
A_\mathrm{C} &= k^2c^2-\frac{\omega^2_{pi}u^2_{di}}{v_{thi}^2}+\omega_{pe}^2, \nonumber\\
C_\mathrm{C} &= 1-\frac{\omega^2_{pi}}{v_{thi}^2k^2}
-\frac{\omega^2_{pi}u_{di}^2}{v_{thi}^4k^2}, \nonumber\\
D_\mathrm{C} &= \omega^2_{pe}v^2_{the}k^2+\omega^2_{pe}u^2_{de}k^2.
\nonumber
\end{align}

\end{widetext}
\end{document}